\documentclass[preprintnumbers]{revtex4}

\usepackage{graphicx,epsfig,rotating}
\def\mpl{\ifmmode \overline M_{Pl}\else $\overline M_{Pl}$\fi}
\def\ra{\rightarrow}
\def\pairof#1{#1^+ #1^-}
\def\ee{\pairof{e}}
\def\ww{W^+W^-}

\def\MPL #1 #2 #3 {Mod. Phys. Lett. {\bf#1},\ #2 (#3)}
\def\NPB #1 #2 #3 {Nucl. Phys. {\bf#1},\ #2 (#3)}
\def\PLB #1 #2 #3 {Phys. Lett. {\bf#1},\ #2 (#3)}
\def\PR #1 #2 #3 {Phys. Rep. {\bf#1},\ #2 (#3)}
\def\PRD #1 #2 #3 {Phys. Rev. {\bf#1},\ #2 (#3)}
\def\PRL #1 #2 #3 {Phys. Rev. Lett. {\bf#1},\ #2 (#3)}
\def\RMP #1 #2 #3 {Rev. Mod. Phys. {\bf#1},\ #2 (#3)}
\def\NIM #1 #2 #3 {Nuc. Inst. Meth. {\bf#1},\ #2 (#3)}
\def\ZPC #1 #2 #3 {Z. Phys. {\bf#1},\ #2 (#3)}
\def\EJPC #1 #2 #3 {E. Phys. J. {\bf#1},\ #2 (#3)}
\def\IJMP #1 #2 #3 {Int. J. Mod. Phys. {\bf#1},\ #2 (#3)}
\def\JHEP #1 #2 #3 {J. High En. Phys. {\bf#1},\ #2 (#3)}
\begin{document}
\bibliographystyle{revtex}

\preprint{SLAC-PUB-9068}

\title{Physics at Multi-TeV Linear Colliders}

\author{Timothy L. Barklow}
\email[]{timb@slac.stanford.edu}
\thanks{Work supported by
Department of Energy contract  DE--AC03--76SF00515.}
\affiliation{Stanford Linear Accelerator Center,
Stanford University, Stanford, California 94309 USA}

\author{Albert De Roeck}
\email[]{Albert.de.Roeck@cern.ch}
\affiliation{CERN, CH-1211, Geneve 23 Switzerland}

\date{December 18, 2001}

\begin{abstract}
The physics at
an $e^+e^-$ linear collider with a center of
mass energy of 3-5 TeV is reviewed.    The following
topics are covered:  experimental environment,
Higgs physics, supersymmetry, fermion pair-production,
$\ww$ scattering, extra dimensions,
non-commutative theories, and black hole production.
\end{abstract}
\maketitle

\section{Introduction}
Presently planned $e^+e^-$ linear collider (LC) projects will operate
at  an initial 
center of mass system (CMS) energy of about 500 GeV, 
with upgrades to higher energies designed in from the start.
The TeV class 
colliders
 TESLA~\cite{tesla} 
and NLC/JLC~\cite{nlc,jlc}  target 800 GeV and 1-1.5 TeV, respectively,
as their maximum CMS energies.
Increasing the energy further would require 
either a change in acceleration technology
or an extension in accelerator length beyond the presently foreseen
 30-40 km~\cite{burrows}. This would also increase the number of active 
elements, which will likely 
decrease the overall efficiency of such a facility.

The nature of the new physics which will hopefully
be discovered  and studied at the LHC and a TeV class LC 
will determine the  
necessity and importance of exploring the multi-TeV range
with a precision machine such as an $\ee$ collider.
This paper summarizes the work of the E3 subgroup 2 on multi-TeV
colliders of the Snowmass 2001 workshop `The Future of Particle Physics'.

Based on our knowledge today, the case for the multi-TeV collider 
rests on the following physics scenarios:

\begin{picture}(200,1)
\put(25,-357){\small \textit{Presented at the APS/DPF/DPB Summer Study on the Future of Particle Physics (Snowmass 2001),}} 
\put(120,-369){\small \textit{30 June - 21 July 2001, Snowmass, Colorado, USA}}
\end{picture}

\begin{itemize}
\item The study of the Higgs\\
For a light  Higgs, a multi-TeV $\ee$ collider can access with high precision
the triple Higgs coupling, providing experimenters with the opportunity to measure the Higgs potential. The 
large event statistics 
will allow physicists to measure rare Higgs decays such as $H\rightarrow
\mu\mu$. For heavy Higgses, predicted by e.g. supersymmetric models, 
the range for discovery and measurement will be extended for masses
up to and  beyond 1 TeV.

\item Supersymmetry\\
In many SUSY 
scenarios   only a subset of the new sparticles will 
be light enough to be produced directly at
a TeV class LC.
Some of the heavier sparticles will be discovered at the LHC, but 
a multi-TeV LC will be needed to complete the spectrum and to
precisely  measure the heavy sparticles properties 
(flavor, mass, width,
couplings). Furthermore, polarized beams  will help 
disentangle mixing parameters and aid CP studies.
Ultimately we {\it will}  need to measure all sparticles 
as precisely as possible to fully pin down and test the underlying theory.

\item New resonances\\
Many alternative theories and models for new physics predict
new heavy resonances with masses larger than 1 TeV.
If these new resonances 
(e.g. new gauge bosons, Kaluza-Klein resonances, or   $W_LW_L$
resonances) have masses
 in the 1~TeV - 5~TeV range, a multi-TeV collider becomes
a particle factory, similar to LEP for the $Z$.
The new particles can be produced
directly  and
their properties can be  accurately determined.

\item No new particles\\
 If {\it no} new particles are observed directly, apart from perhaps 
one light Higgs particle, then a 
multi-TeV collider will probe new physics
indirectly (extra dimensions, $Z'$, contact interactions) at scales in the
range of
30-400 TeV 
via precision
measurements.

\item Unexpected phenomena \\
This is probably the most exciting of all: perhaps Nature 
has chosen a road
as yet not explored (extensively) by our imagination. Recent  examples 
of new ideas are 
 string quantum gravity effects, non-commutative effects, black hole
formation, nylons, and split fermions.

\end{itemize}

\begin{figure}[htb]
\epsfig{file=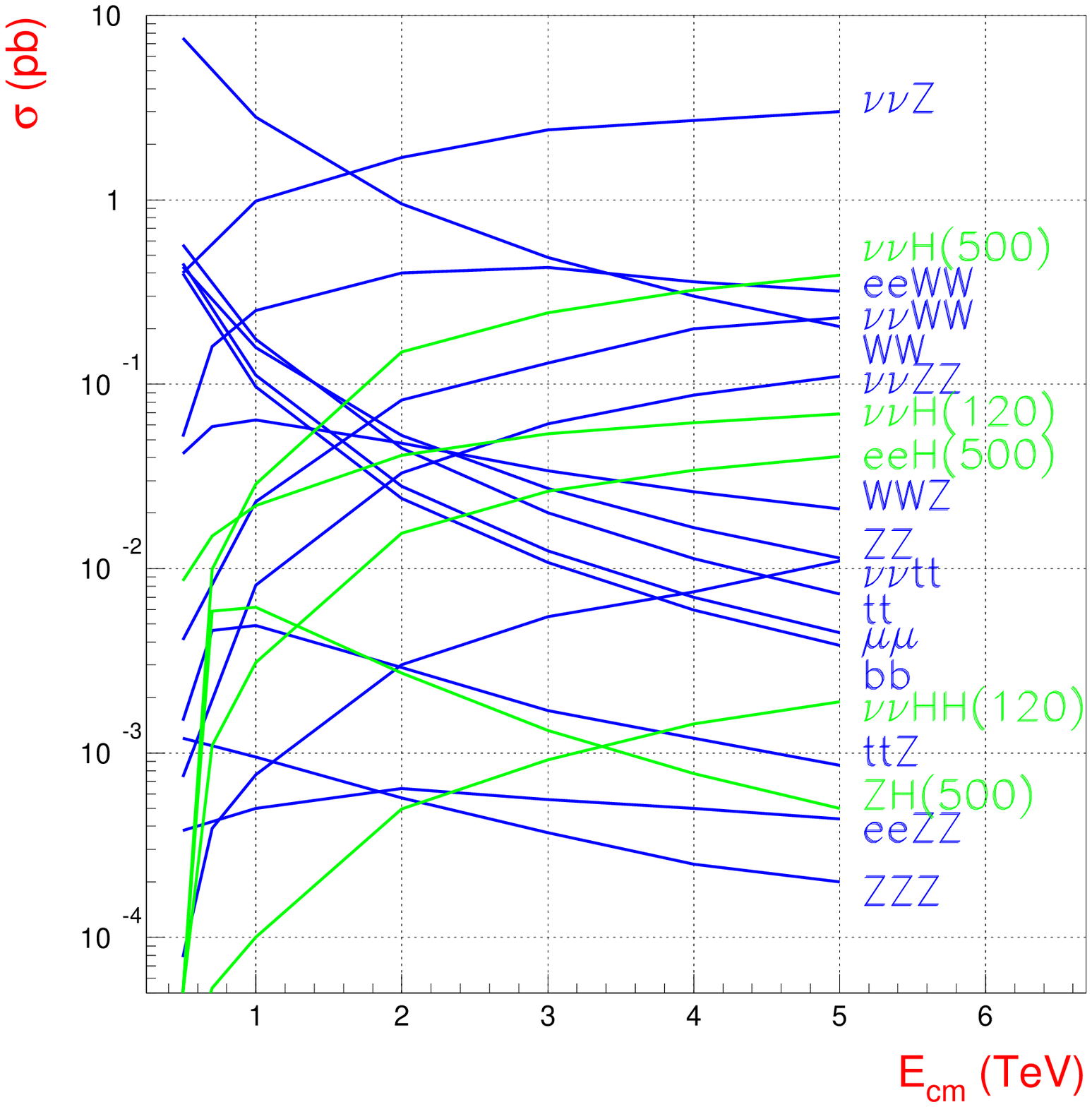,bbllx=15,bblly=15,bburx=520,bbury=520,width=8cm}
\epsfig{file=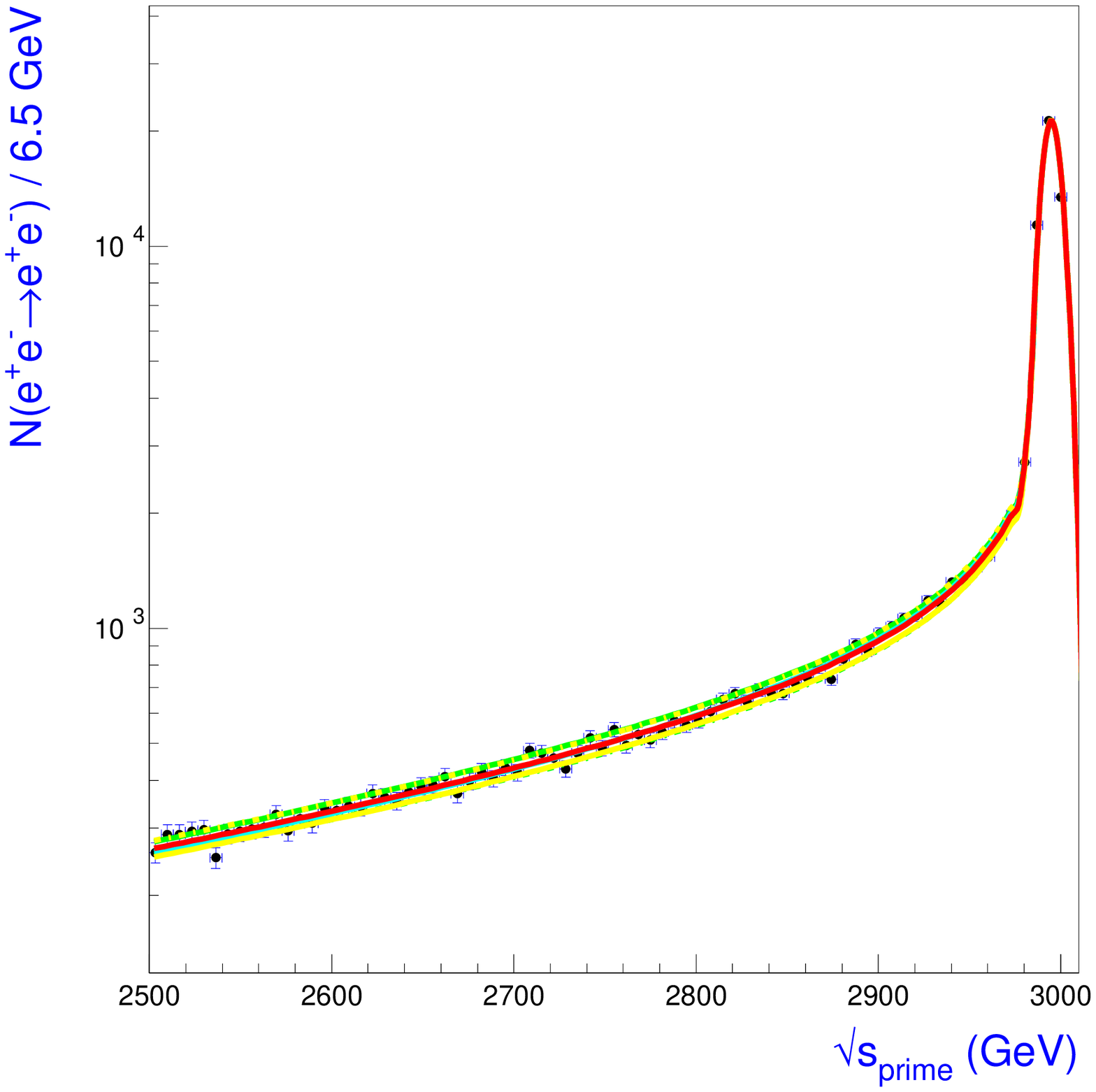,bbllx=15,bblly=10,bburx=520,bbury=520,width=8cm}
\caption{(Left) Cross sections for several $s$ and $t$ channel
exchange processes.
(Right) Example of a luminosity spectrum at CLIC~\protect \cite{baggzp}.}
\label{fig:cross1}
\end{figure}

\begin{table}[htbp]
\caption{Event rates for several processes in the multi-TeV
range, for 1 ab$^{-1}$ integrated luminosity.}
\begin{center}
\begin{tabular}{|c|c|c|}
\hline
Event Rates/Year  &  3 TeV & 5 TeV \\
(1000 fb$^{-1}$) &  $10^{3}$ events  & $10^{3}$ events\\
\hline
$e^+e^-\rightarrow t\overline{t}$  & 20 & 7.3 \\
$e^+e^-\rightarrow b\overline{b}$  & 11 & 3.8 \\
$e^+e^-\rightarrow ZZ$  & 27 & 11 \\
$e^+e^-\rightarrow WW$  & 490 & 205 \\
$e^+e^-\rightarrow hZ/ h\nu\nu$ (120 GeV)  & 1.4/530  & 0.5/690 \\
$e^+e^-\rightarrow H^+H^-$(1 TeV)  & 1.5 & 0.95 \\
$e^+e^-\rightarrow \tilde{\mu}^+\tilde{\mu}^-$ (1 TeV)& 1.3 & 1.0 \\
\hline
\end{tabular}
\label{tab:tab1}
\end{center}
\end{table}

An increase of CMS energy  needs to be accompanied
 by an increase
of luminosity to compensate for the $1/s$ dependence of the 
$s$-channel annihilation
cross section. However,  a feature of the multi-TeV energy range is that the
fusion
processes ($t$-channel exchange), which increase logarithmically with 
CMS energy, become 
comparable in strength to the $s$-channel annihilation processes,
as can be seen in Fig.~\ref{fig:cross1}.
The event rate of several processes at 3 and 5 TeV 
CMS energy is given in Table~\ref{tab:tab1}~\cite{baggzp}.

\begin{figure}[htbp]
\epsfig{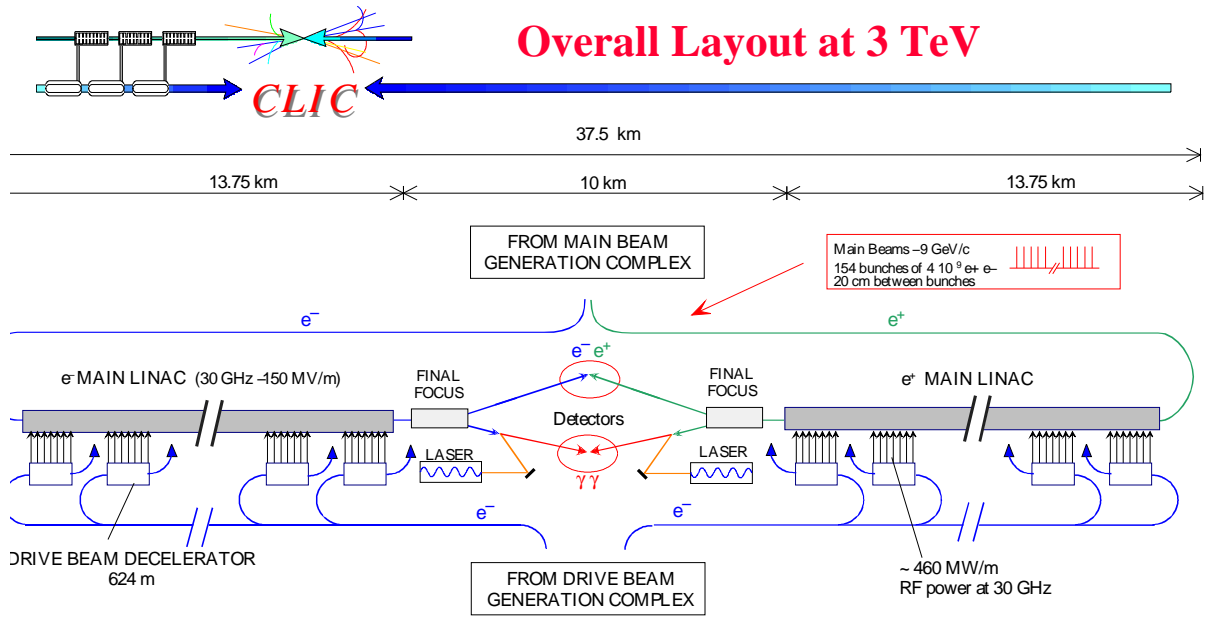}
\caption{Overall layout of the CLIC complex for a center-of-mass energy
of 3 TeV.}
\label{fig:clic}
\end{figure}


The CLIC project at CERN is studying  the feasibility of  
an $e^+e^-$ linear collider optimized for 
a CMS energy of  3~TeV with 
 ${\cal L} \cong 
10^{35}$cm$^{-2}$s$^{-1}$, using
a novel  technique called two-beam acceleration~\cite{clic}.
A so called drive beam of low energy but high current
is decelerated, and its energy is transferred to the main low
current beam, which 
 gets accelerated with gradients in the range of 100-200 MV/m.
The layout of a CLIC accelerator is shown in Fig.~\ref{fig:clic}.
With an upgrade program a maximum energy of 5 TeV is foreseen.

In order to reach such high luminosities CLIC will operate in the 
high beamstrahlungs regime, and the 
 beam parameters and machine requirements are very challenging: nm stability
of the components,
strong final focus, and 30 GHz accelerating structures.
Increasing the luminosity beyond $10^{35}$cm$^{-2}$s$^{-1}$ 
will certainly be hard, and probably can only be done by increasing 
the power and the number of bunches. Likewise, an increase of the 
energy, e.g. towards 10 TeV, is not excluded but so far only via extending the 
length of the accelerator.

Several test facilities (CTF1, CTF2)~\cite{ctf} 
have been built over the past 5 years which
have succeeded in demonstrating the principle of two beam acceleration.
So far, in   CTF2 one has reached gradients
of 72 MV/m (15 ns pulses, gradient measured with the beam), 
and 160 MV/m (3 ns pulses, value determined from the power).
Furthermore, an  X-band structure has been 
built which achieved 150MV/m peak accelerating field 
without beamloading,
for 150ns.

A new test facility, CTF3~\cite{ctf3}, is proposed for  2002-2006 and
should test the drive beam generation concept.
With the knowledge gained at this test facility, if no bad surprises emerge, 
a conceptual design report for CLIC can be completed in the years 
afterwards. When it has been decided to develop a full  CLIC facility, it
is likely to start off with a lower  energy version, e.g. 
with a Higgs factory in $\gamma\gamma$ mode
as discussed in~\cite{velasco}, to gain experience with two beam
acceleration before making  the big leap to the 
multi-TeV region.

A scheme  for an adiabatic upgrade
of the JLC/NLC into the multi-TeV region has also been 
presented~\cite{burrows}.   Through the phased installation of a low energy, high current drive beam 
in the JLC/NLC tunnel, the JLC/NLC could be gradually transformed from an
X-band klystron powered accelerator into a 
two beam 
accelerator \`a la CLIC.

A multi-TeV collider will have all the features of a TeV class 
linear collider, 
including $e^-e^-$, $e\gamma$ and $\gamma\gamma$ collider options.
Realizing these options in the multi-TeV region is somewhat more difficult
than for a sub-TeV collider, 
but first studies indicate that the additional complications
can probably be overcome~\cite{telnov}.
Polarized beam collisions 
 will have an additional  loss of 7\% of effective polarization 
 during the beam-beam 
interaction, due to the beamstrahlung and strong fields at 
3 TeV~\cite{assman}.
Therefore,  polarimeters before and 
after the collision point will be  needed to control this effect.
Physics processes might also be used to monitor the effective 
polarization at the interaction point (IP).

\section{Experimental Environment}

The machine parameters~\cite{clic,guignard} needed to achieve these energy and 
luminosity values 
 lead to important challenges for experiments at CLIC. The 
beam-beam effects result in  considerable backgrounds and a distortion
of the luminosity spectrum.
An example of a luminosity spectrum is shown in Fig.~\ref{fig:cross1}.
The effective luminosity available in the peak is given in 
Table~\ref{tab:tab3} for different CLIC CMS energies.
A list of parameters and backgrounds 
which will determine the experimental environment
 at CLIC is given in
Table~\ref{tab:tab2}
 for 3 TeV and ${\cal L} = 10^{35}$cm$^{-2}$s$^{-1}$.

The total number of produced $e^+e^-$ pairs 
is huge; however the
 coherent pairs disappear essentially down the beampipe, and
the  incoherent pairs can be  suppressed almost completely if 
 the detector includes a  strong magnetic field. These backgrounds force the 
inner vertex detector to have a minimal inner radius of 3cm. 
There are additional backgrounds which need to be taken into account,
 such as
neutrons, muons, and synchrotron radiation, and these 
are presently  under study~\cite{schulte}.

\begin{table}
\begin{center}
\caption{Luminosity within 1\% \& 5\% of $2\cdot E_{beam}$ at CLIC}
\begin{tabular}{|c|c|c|c|c|}
\hline
Energy (TeV) & 0.5 & 1 & 3 & 5 \\
\hline
${\cal L}$ in 1\% \boldmath $\sqrt{s}$ & 71\%  & 56 \% & 30\% & 25\% \\
\hline
${\cal L}$ in 5\% \boldmath $\sqrt{s}$ & 87\%  & 71 \% & 42\% & 34\% \\
\hline
\end{tabular}
\label{tab:tab3}
\end{center}
\end{table}

\begin{table}
\caption{Parameters for experimenting at CLIC,
 for 3 TeV and ${\cal L} = 
10^{35}$cm$^{-2}$s$^{-1}$.}
\begin{center}
\begin{tabular}{|c|c|c|}

\hline
 Luminosity/bunch & $10^{-2}$ nb$^{-1}$\\
Beam energy spread & $1\% $(FWHM)\\
bunches per train/ Rep. rate &  $154/100$ \\
Time between bunches &  $0.67$ ns \\
 Average energy loss & $31$\% \\
 photons/beam particle &   $2.3$ \\
 Number/energy incoh. pairs &
  $4.6\cdot 10^5/3.9\cdot 10^4$ TeV \\
 Number/energy coh. pairs &  
  $1.4\cdot 10^9/ 4.4 \cdot 10^{8}$ TeV \\
  Hadronic ($\gamma\gamma$) events, $W_{\gamma\gamma}>5 $ GeV &    
 $4$\\
 \hline
\end{tabular}
\label{tab:tab2}
\end{center}
\end{table}

Based on the above machine parameters,  the experience at LEP/SLC, and 
the developments made for TESLA/NLC~\cite{settles},   
a first layout for a detector
has been worked out. 
So far the emphasis  has  been on
 tracking resolution,
jet flavor tagging, energy flow, hermeticity,
 precise vertexing, and high calorimeter granularity.
First ideas for a detector layout at CLIC
are given in Table~\ref{tab:tab4}, based on a 4T magnetic field.
A detector with a 5-6T field, which could be more compact, is considered
as well.

\begin{table}
\begin{center}
\caption{Design concept (detector layers) for a detector at CLIC}
\begin{tabular}{|l|l|}
\hline
3--15 cm & Silicon VDET \\
15--80 cm & Silicon central/forward disks\\
80--230 cm & TPC or Silicon tracker \\
240--280 cm & ECAL (30 $X_0$)\\
280--400 cm & HCAL (6$\lambda$)\\
400--450 cm & Coil (4T)\\
450--800 cm & Fe/muon\\
\hline
\end{tabular}
\label{tab:tab4}
\end{center}
\end{table}
 
However, it is not clear that these lower energy
concepts will still be usable at CLIC energies. 
The multi-TeV high energy frontier opens up new questions such as
 timing issues ( $< 1$~ns distance between bunches),  energy flow
(many highly collimated jets) ,
and displaced vertex flavor tagging.
The latter is demonstrated in Table~\ref{tab:tab5}: at 
$\sqrt{s}$ = 3 TeV the $B$ mesons produced in $b\bar{b}$ events
decay on average a
distance of  9.0~cm from the primary vertex, hence they will decay
mostly within or after 
the vertex detector. For the CLIC physics
studies a technique based on multiplicity
increase versus distance from the vertex has been 
developed~\cite{battvtx} to search
for $B$-decays. 
It has been shown that this technique can tag $B$ mesons with
an efficiency larger than 50\% and purity of 85\% per jet.

\begin{table}
\begin{center}
\caption{B-decay length in space at different CMS energies, and for 
several processes.}
\begin{tabular}{|l|c|c|c|c|c|}
\hline
$\sqrt{s}$ (TeV) & 0.09  & 0.2 & 0.35 & 0.5 &   3.0 \\ \hline
                 & $Z$ & $HZ$ & $HZ$ & $HZ$ & $H^+H^-$ $|$ 
 $b \bar b$ \\
\hline 
$d_{space}$ (cm) & 0.3 & 0.3 & 0.7 & 0.85 & ~~~  2.5~~~~$|$ 
  9.0\\
\hline
\end{tabular}
\label{tab:tab5}
\end{center}
\end{table}
Just as for TeV class linear 
colliders, the background in the interaction region
 at CLIC requires the use of a mask. For CLIC
the mask covers  the region below 120 mrad; 
electron/photon
 tagging will be possible down to 40 mrad with detectors in the mask.

The CLIC physics working group has developed 
a set of tools to enable physicists to rapidly 
study physics processes within the multi-TeV experimental environment.
 The CLIC luminosity spectrum is generated with CALYPSO~\cite{schulte}.
 The hadronic background from
$\gamma\gamma$ events (4 events/bunch crossing) is generated with HADES~\cite{schulte}.
 The SIMDET-CLIC fast simulation package is used (based on the 
package for the TESLA Detector~\cite{simdet}).
 The adapted VECSUB package for jet reconstruction is used 
(DURHAM, JADE algorithms)~\cite{vecsub} for analysis.
Most of the studies discussed in this paper are based on these tools.

\begin{figure}[htbp]
\begin{center}
\epsfig{file=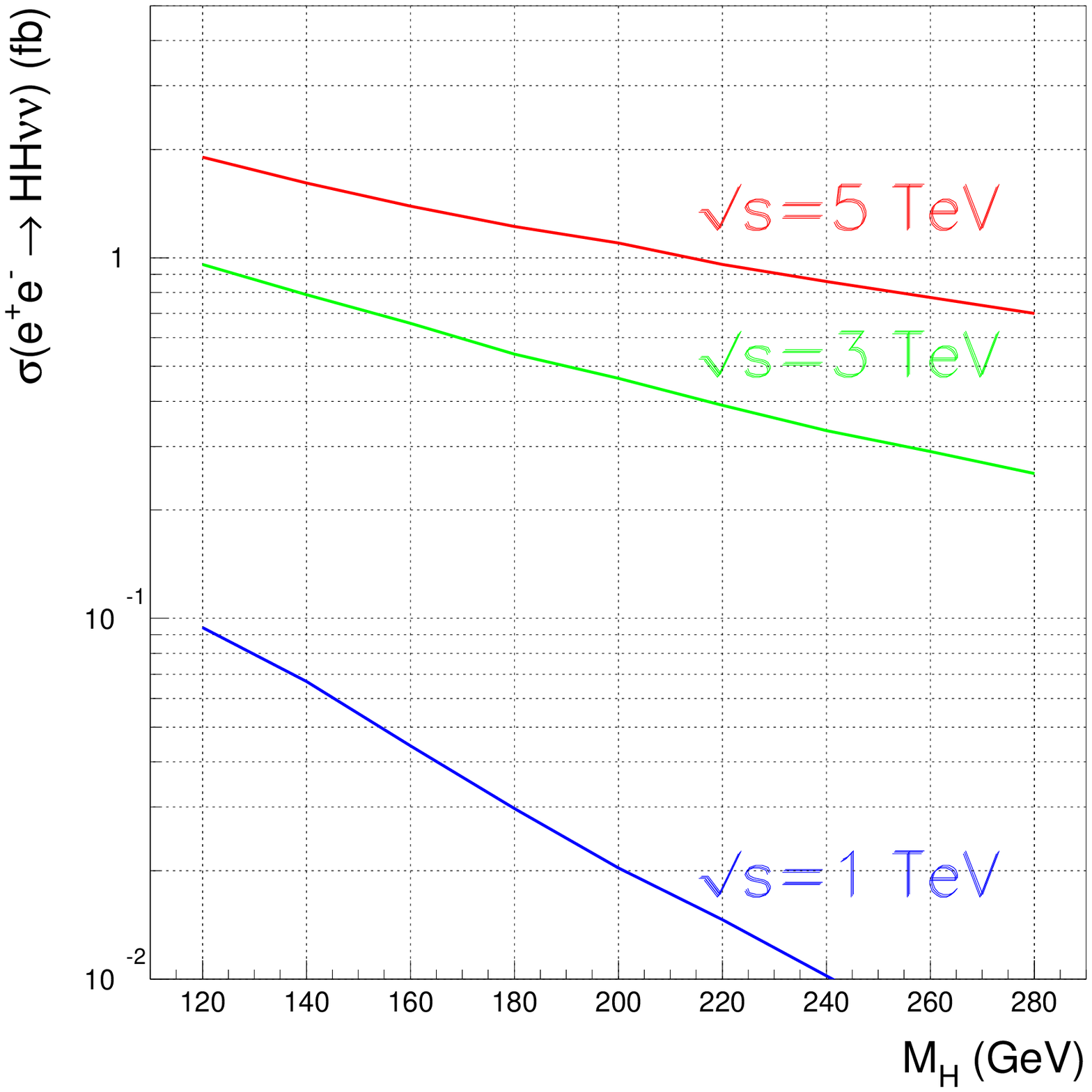,bbllx=15,bblly=15,bburx=520,bbury=520,width=8cm}
\epsfig{file=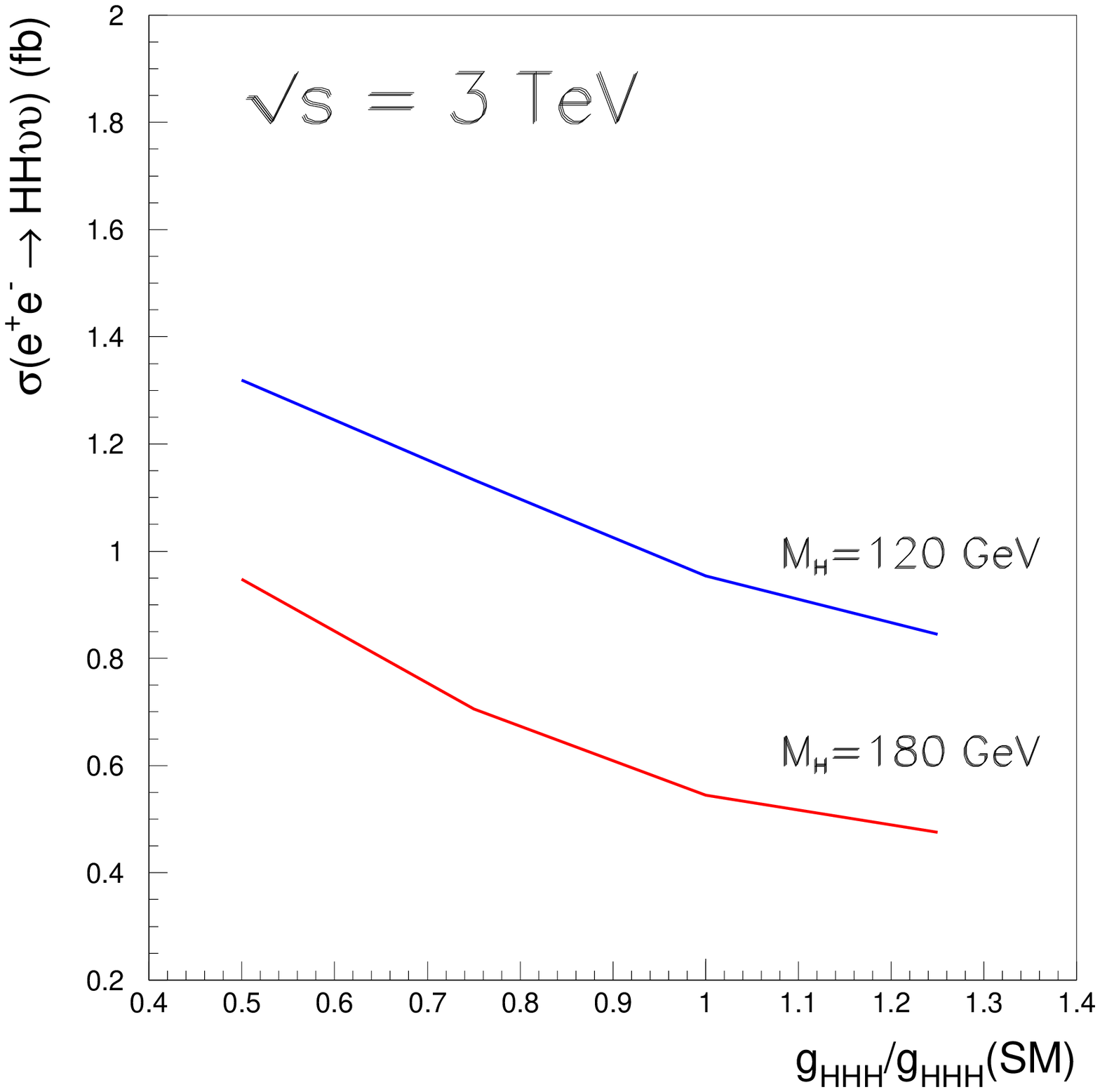,bbllx=15,bblly=15,bburx=520,bbury=520,width=8cm}
\caption{(Left)
$\sigma_{HH\nu\nu} $
as function of $M_H$ for  a LC with 
$\sqrt{s} = 0.5, 1$ and 3 TeV. (Right) Change of the cross section 
for $e^+e^- \rightarrow HH\nu\nu$ as function of the change 
of the coupling $g_{HHH}$ \protect \cite{batthiggs1}.}
\label{fig:hnn}
\end{center}
\end{figure}

\section{Higgs Physics}


Within the SM the circumstantial evidence for a light Higgs is 
accumulating, offering the TeV class LCs  an exciting physics program
to study in detail the Higgs properties. However -- perhaps at first 
sight surprisingly -- a multi-TeV can  add substantial 
and important information about a light Higgs.

A final proof of the Higgs mechanism, which may still be  unsettled
after the TeV class LC data has been collected, is the
accurate reconstruction 
of the potential of the Higgs field. 
The Higgs potential $V$ can be studied through triple ($g_{HHH}$) and quartic 
Higgs couplings:
\begin{equation}
V = \lambda v^2 H^2 + \lambda v H^3 + \textstyle{\frac{1}{4}} \lambda H^4,
\,\,\,\,\,\,  M_H=\sqrt{2\lambda} v, \,\,\,\,\,\,\, g_{HHH} = 3\lambda v\ , 
\end{equation}
where $H$ is the Higgs field, $M_H$ is the Higgs mass and $v=246$~GeV is the Higgs vacuum expectation value parameter.
An accurate measurement of this relation would probe the extended 
nature of the Higgs sector, and is therefore of fundamental importance.

The triple Higgs coupling can be studied in double Higgsstrahlung,
$e^+e^-\rightarrow HHZ$, and in the fusion process 
$e^+e^- \rightarrow HH\nu\nu$. Only the second process is relevant for 
multi-TeV colliders. The cross section as a function of the Higgs mass is shown
in Fig.~\ref{fig:hnn}
for several LC CMS energy values: the cross section is very strongly 
CMS energy dependent. Together with the larger luminosity expected for 
a multi-TeV collider, this will lead to
 event samples  which are up to a 
factor of 100 larger than at a TeV class collider.

A study including detector acceptance, smearing and background 
was performed for 
$4b$ ($M_H$ = 120 GeV) and
$4W$ ($M_H$ = 180 GeV) final states~\cite{batthiggs1}.
The statistical precision on $g_{HHH}$
with 5 ab$^{-1}$ and a CMS energy of
3 TeV from a measurement of the 
$HH\nu\nu$ cross section for
$M_H = 120 (180)$ GeV is
$\delta g_{HHH}/g_{HHH} = 0.094 (0.140)$. 
By making use of the scalar nature of
the Higgs decay these 
limits can be improved to
$\delta g_{HHH}/g_{HHH} = 0.070 (0.080)$ for
$M_H = 120 (180)$ GeV.
Clearly this channel will be statistics limited. 
The use of polarized beams could yield up to 
a factor of about 4 more in statistics.
No other machine presently considered could reach this kind of precision
in the triple Higgs coupling.
Fig.~\ref{fig:hnn} shows the dependence of the $HH\nu\nu$ cross section
on the triple Higgs coupling, normalized to its SM value, for 
$\sqrt{s} = 3$~TeV and two values of $M_H$.

The benefit of an increasing event rate  with increasing
 CMS energy does not continue indefinitely, due to the 
competition of background processes which have  the same final state but 
no triple Higgs vertex, and due to the forward nature of the 
produced Higgs pairs, which become increasingly more difficult to detect.
Studies indicate that the optimal CMS energy for a 
maximum sensitivity to the triple Higgs coupling is between 2.5 and 4 TeV,
depending on the Higgs mass.

The quartic Higgs coupling remains 
unfortunately elusive. The cross section for
$ee\rightarrow HHH\nu\nu$ is about
three orders of magnitude smaller than for
$ee\rightarrow 
HH\nu\nu$. Hence,
at a 3 TeV collider one expects only 0.4 events produced per year.
In the most favorable condition of a 10 TeV
collider and $10^{35}$cm$^{-2}$s$^{-1}$, one would produce only 5 
signal events
per year, so that even the prospect for exploratory studies looks rather dim.

\begin{figure}
\epsfig{file=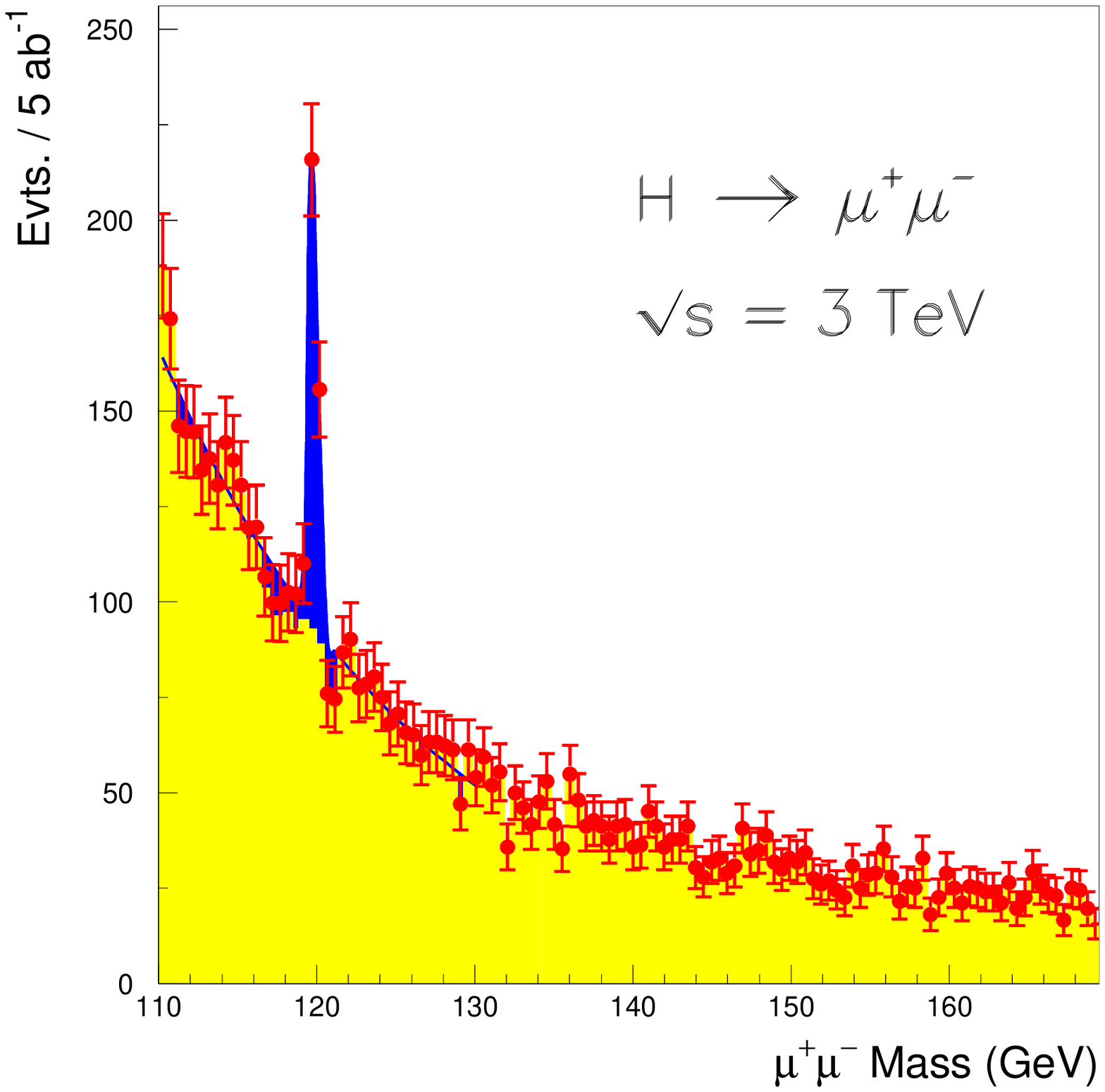,bbllx=0,bblly=0,bburx=580,bbury=600,width=5cm}
\epsfig{file=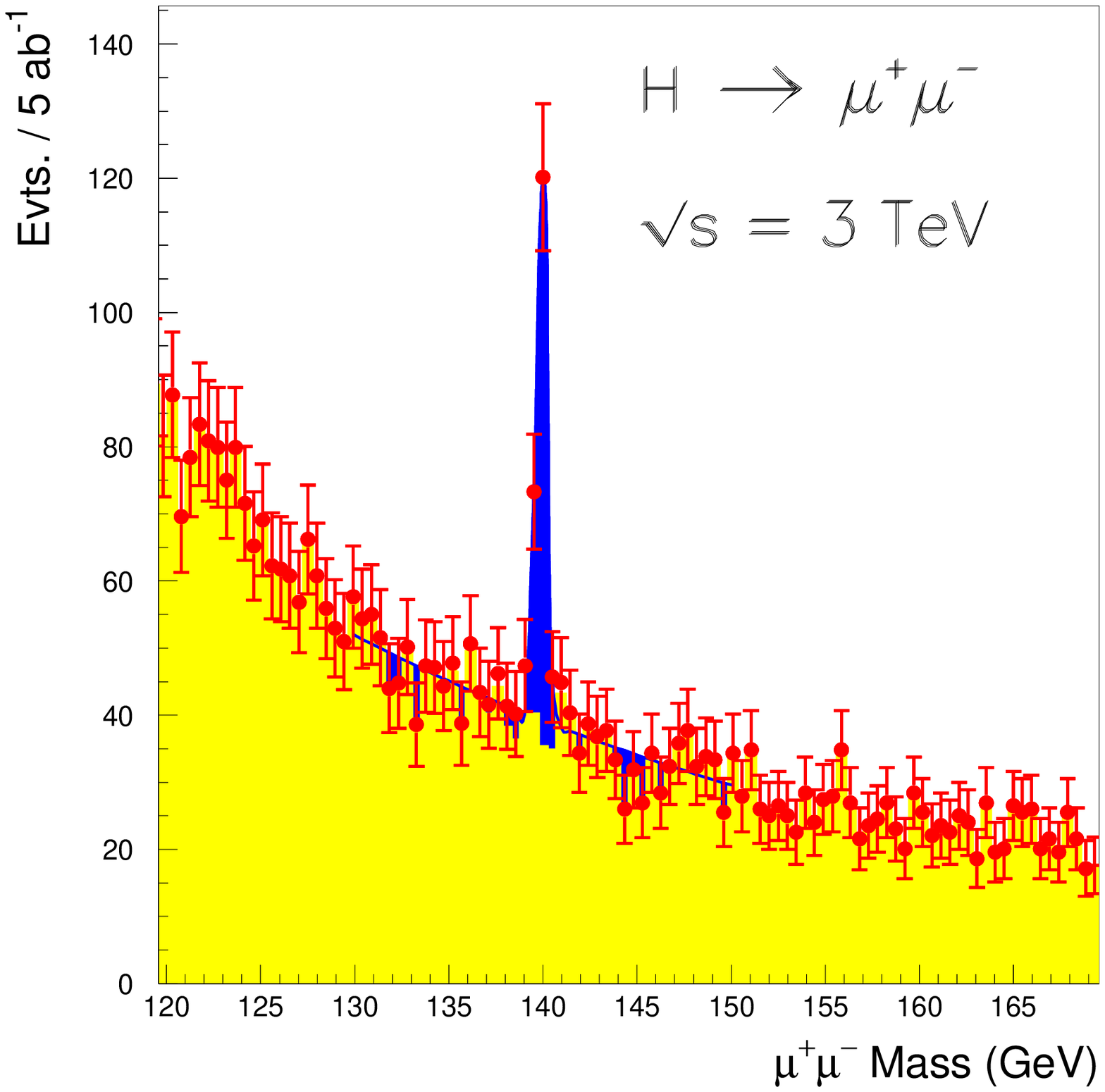,bbllx=0,bblly=0,bburx=580,bbury=600,width=5cm}
\epsfig{file=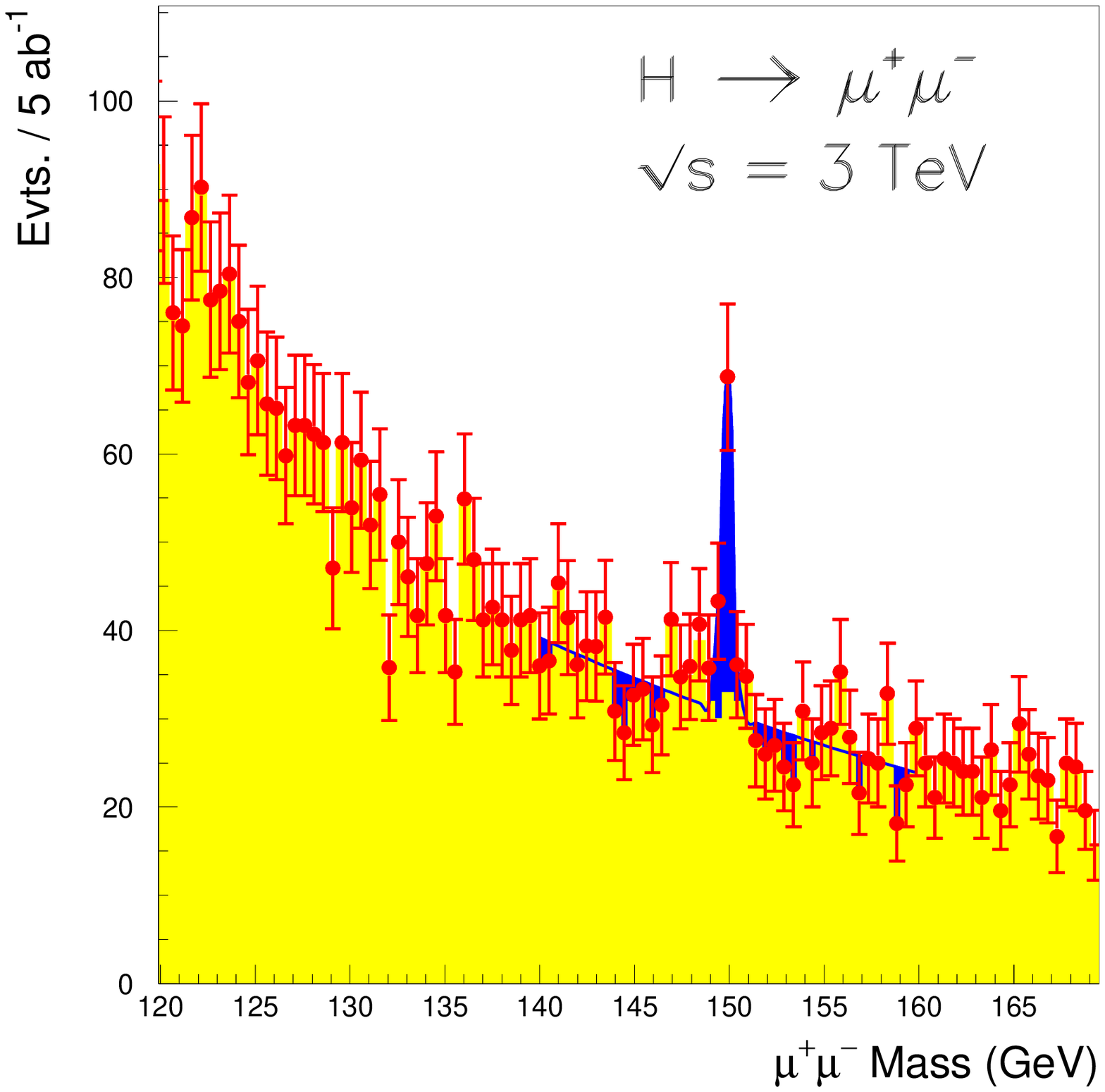,bbllx=0,bblly=0,bburx=580,bbury=600,width=5cm}
\caption{The $\mu\mu$ invariant mass spectrum for background and 
signal $H\rightarrow \mu\mu$ for different values of 
$M_H$~\protect \cite{battaglmumu}.}
\label{fig:mumu} 
\end{figure}

Apart from the Higgs potential, a multi-TeV collider also provides physicists with the opportunity to 
measure rare decay modes of the Higgs.  For example, the 
decay $H\rightarrow \mu\mu$ has a branching ratio of about $10^{-4}$
in the SM for a light Higgs. Its measurement 
is  one of the driving motivations for a muon collider.
For 5 ab$^{-1}$ at 3 TeV, $2.7\times 10^6$ Higgs bosons with $M_H$ = 120 GeV
will be produced, of which 650 will decay into muons.
Including detector smearing and acceptance leads to  a di-muon 
invariant mass spectrum as shown in Fig.~\ref{fig:mumu}~\cite{battaglmumu}.
The measurement of this cross section can be combined with an independent
measurement of the Higgs
production cross section, e.g. via the $b$ or $W$
decay modes, to extract $g_{H\mu\mu}$.
The statistical accuracy of the measurement of  the coupling 
$g_{H\mu\mu}$ can be 
$\delta g_{H\mu\mu}/g_{H\mu\mu} = $ 0.040, 
0.062, and 0.106, for Higgs masses of 120, 140, and 150 GeV, respectively. Hence the anticipated accuracy 
of this measurement
is comparable to that predicted at a muon collider~\cite{muoncol}, and better 
than present expectations for a VLHC at 200 TeV~\cite{bauer}.
Note that for this channel the production is rather forward, so that 
a further increase in CMS energy does not translate completely
to a  corresponding
increase in precision.

With the measurement of this branching ratio 
one can verify  the Higgs mechanism in the lepton  sector -- specifically
the relation $g_{H\mu\mu}/g_{H\tau\tau} = M_{\mu}/M_{\tau}$ --
with a precision of 5-8\% for a Higgs mass in the range of 
120-140 GeV. Any striking anomalous behaviour in the muon channel would 
be easily detectable.

In  minimal SUSY scenarios  a  total of 5 Higgs particles is  expected.
One light neutral higgs is expected, typically with a mass below 130 GeV,
while the masses of the others Higgses 
are usually degenerate and depend strongly on the SUSY parameters 
of the model.
Scenarios have emerged where the heavy Higgses are larger than 500 GeV, and 
therefore cannot be pair produced at a 1 TeV  LC. Some gain in reach
can be obtained, up to about 
$M_H = 700$ GeV, using the LC in the $\gamma\gamma$
mode. 

\begin{figure}
\begin{center}
\epsfig{file=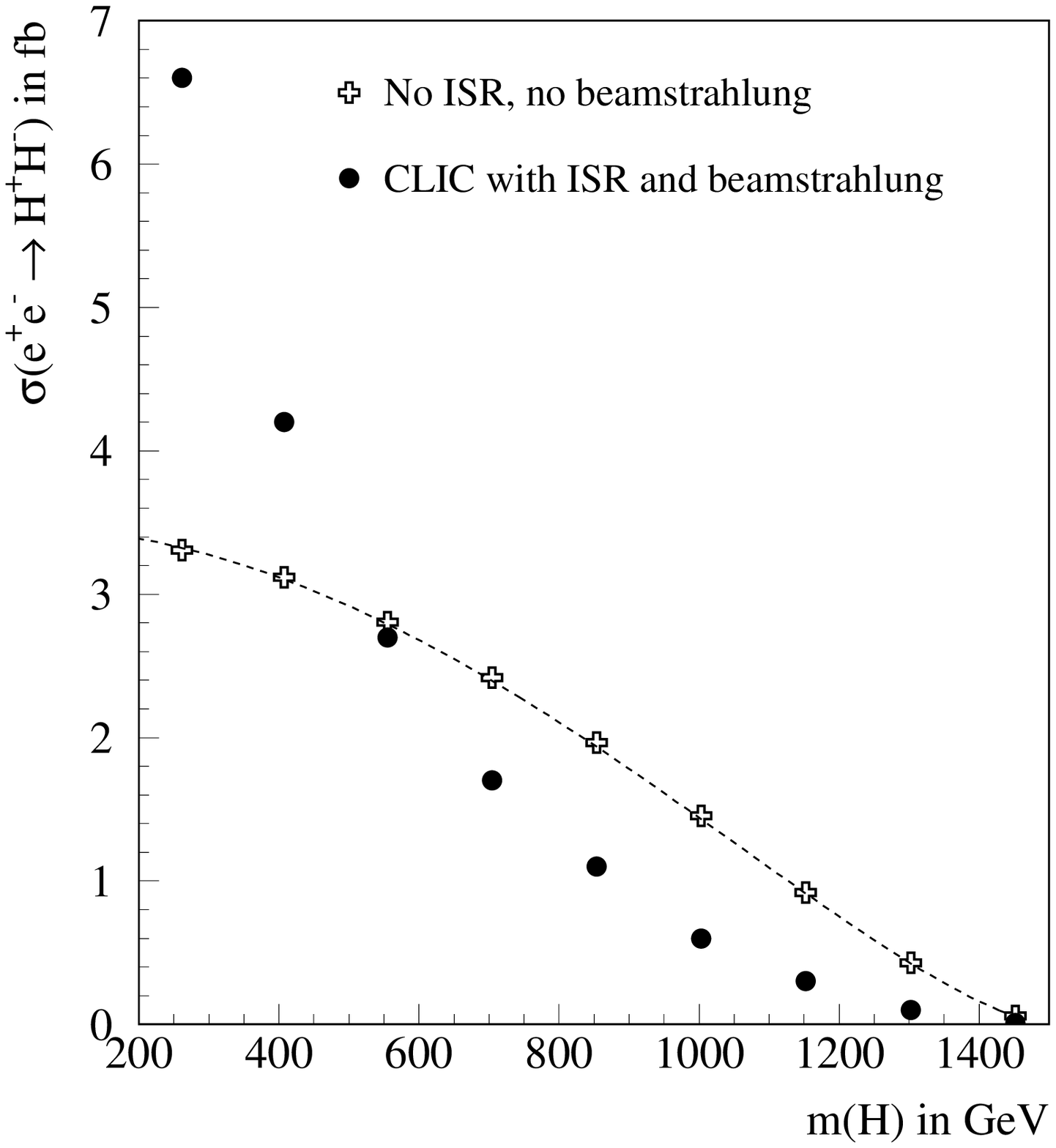,bbllx=0,bblly=0,bburx=460,bbury=500,width=8cm,clip=}
\epsfig{file=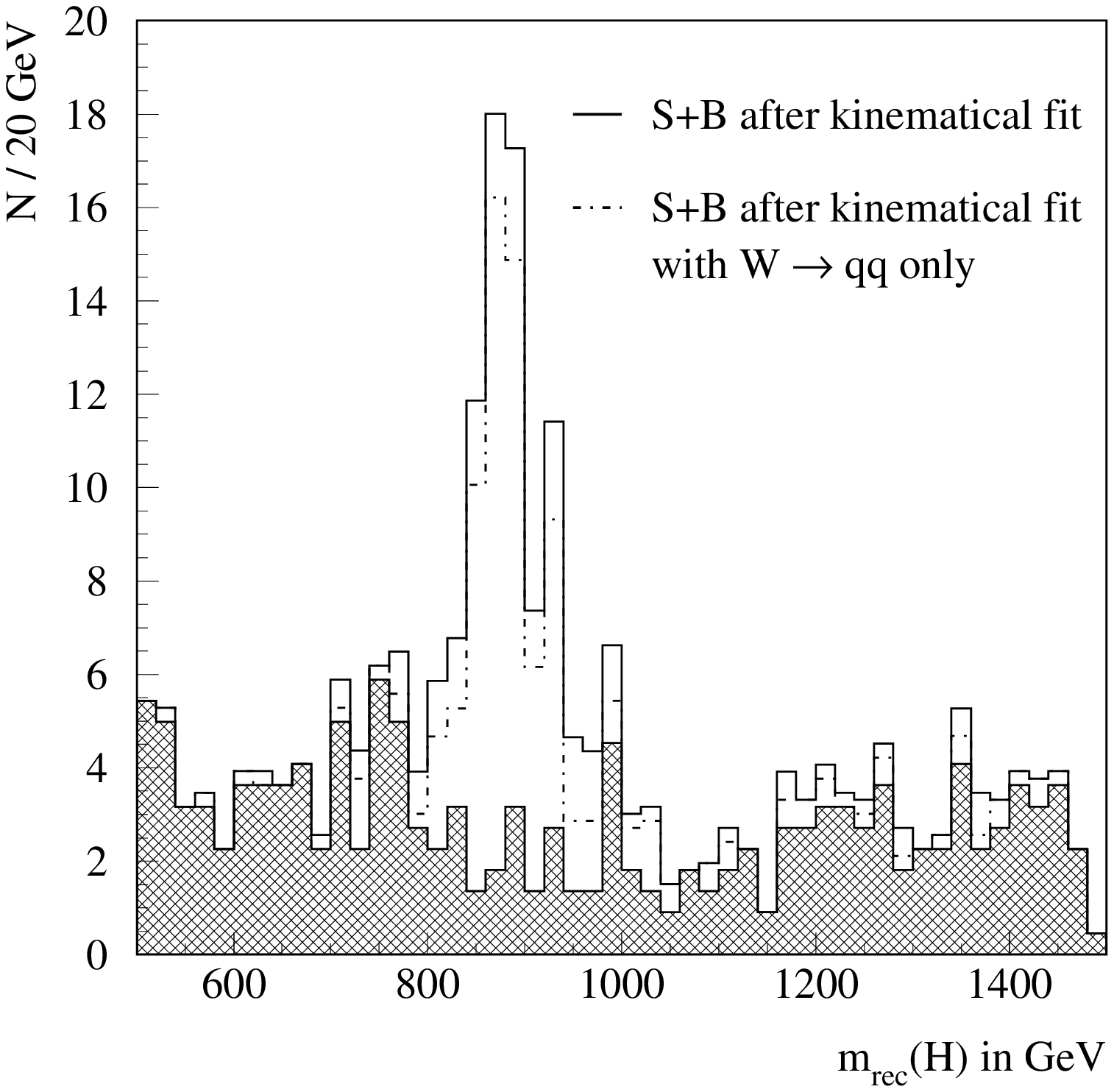,bbllx=0,bblly=0,bburx=460,bbury=500,width=8cm,clip=}
\end{center}
\caption{(Left) Cross section for $e^{+}e^{-} \rightarrow H^{+}H^{-}$ at 
  $\sqrt{s}=3~\mbox{TeV}$. 
Open crosses and dashed lines show the cross section 
at tree level, while full circles show the effective cross 
section after including ISR and beamstrahlung.
(Right) Reconstructed $H^{\pm} \rightarrow tb$ invariant 
mass after the kinematical fit
 for 
an integrated luminosity of 3~ab$^{-1}$. The 
$t\overline{b}\overline{t}b$ background 
is shown by the hatched distribution\protect \cite{ferrari}.}
\label{fig:heavyhiggs}
\end{figure}

For CLIC the process    
$e^+e^-\rightarrow H^+H^-$ with $M_{H} = 880$ GeV has been studied.
A scenario including heavy Higgs masses around this value
is given by point (J) in the newly 
proposed SUSY benchmarks~\cite{ellis}, discussed in the next section.
The decay channel $H^{\pm} \rightarrow tb \rightarrow Wbb \rightarrow qqbb$
has been studied in which eight jets are expected in the final 
state~\cite{ferrari}. Events are selected where both
 Higgses are reconstructed.
The tagging of $b$-quarks  will be extremely important for such an  analysis.
The cross section is shown in 
Fig.~\ref{fig:heavyhiggs}.
Using the kinematic constraints of the $W$ and $t$ masses to disentangle
the jet pairing, events with two heavy  Higgses
 can be selected. An overall kinematic fit, taking into account 
the effects of the overlaid $\gamma\gamma$ events, can improve
the Higgs mass resolution by a 
factor two, leading to a resolution of 33 GeV which can be compared to 
the  natural width of the $H^{\pm}$ of 21 GeV. 
The result is shown in Fig.~\ref{fig:heavyhiggs}, for a study 
including 15 bunch crossings 
with $\gamma\gamma$ events  overlaid. A total of 47 reconstructed events of the type
$e^+e^- \rightarrow H^+H^-\rightarrow (t\overline{b})(\overline{t}b)$
is  expected in the mass region of 
2.5$\sigma$ around the peak.
From these and follow up 
studies we find that a 5$\sigma$ discovery can be made 
at a 3 TeV collider with 3 ab$^{-1}$ for a Higgs mass up to 
about 1.1 TeV.
Once the heavy Higgs has been 
discovered one can increase the efficiency (now a only few \%)
and hence the number of heavy Higgs particles by using a mass window and 
requiring only (at least) one Higgs per event. 

In short a multi-TeV collider will add new information on the 
measurement of the Higgs potential, rare Higgs decays, and  heavy Higgs states.

\section{Supersymmetry}

A strong candidate for
new physics, which would explain the hierarchy problem, is supersymmetry.
This new symmetry predicts a partner for each known particle,
 a
``sparticle''. Since no sparticles have been observed so far supersymmetry is
broken at the electroweak scale, and there are many models
proposed for
the supersymmetry breaking mechanism. These supersymmetry breaking
models reduce the
more than   
100 new parameters that enter the theory to  only a handful.
All of these models predict a different
phenomenology, including different mass spectra for the
supersymmetric partners.  Sparticles can be readily discovered at an   
$e^+e^-$ machine 
so long as direct production is kinematically allowed.

\begin{figure}
\begin{center}
\epsfig{file=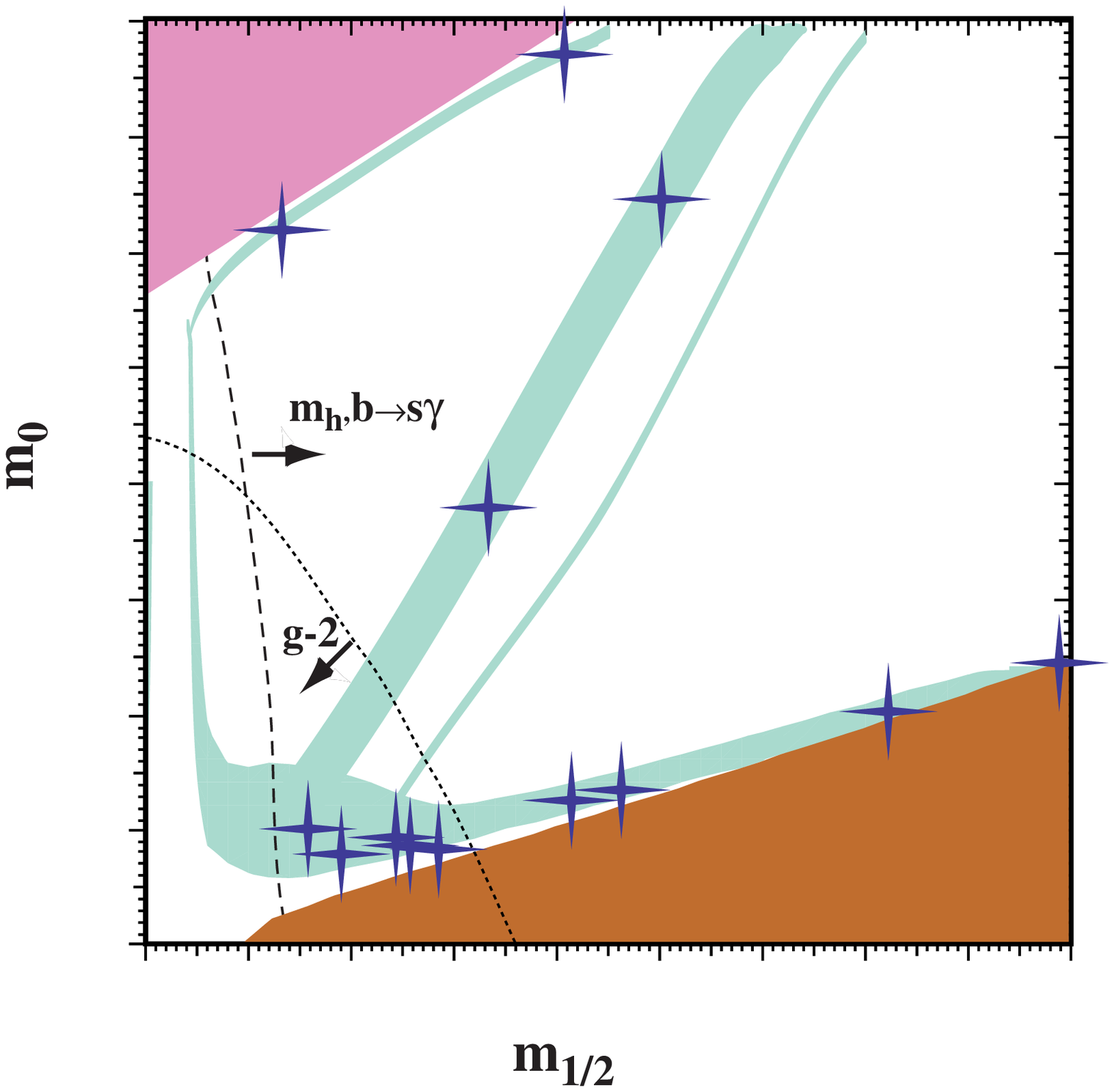,bbllx=105,bblly=175,bburx=550,bbury=640,width=8.0cm}
\epsfig{file=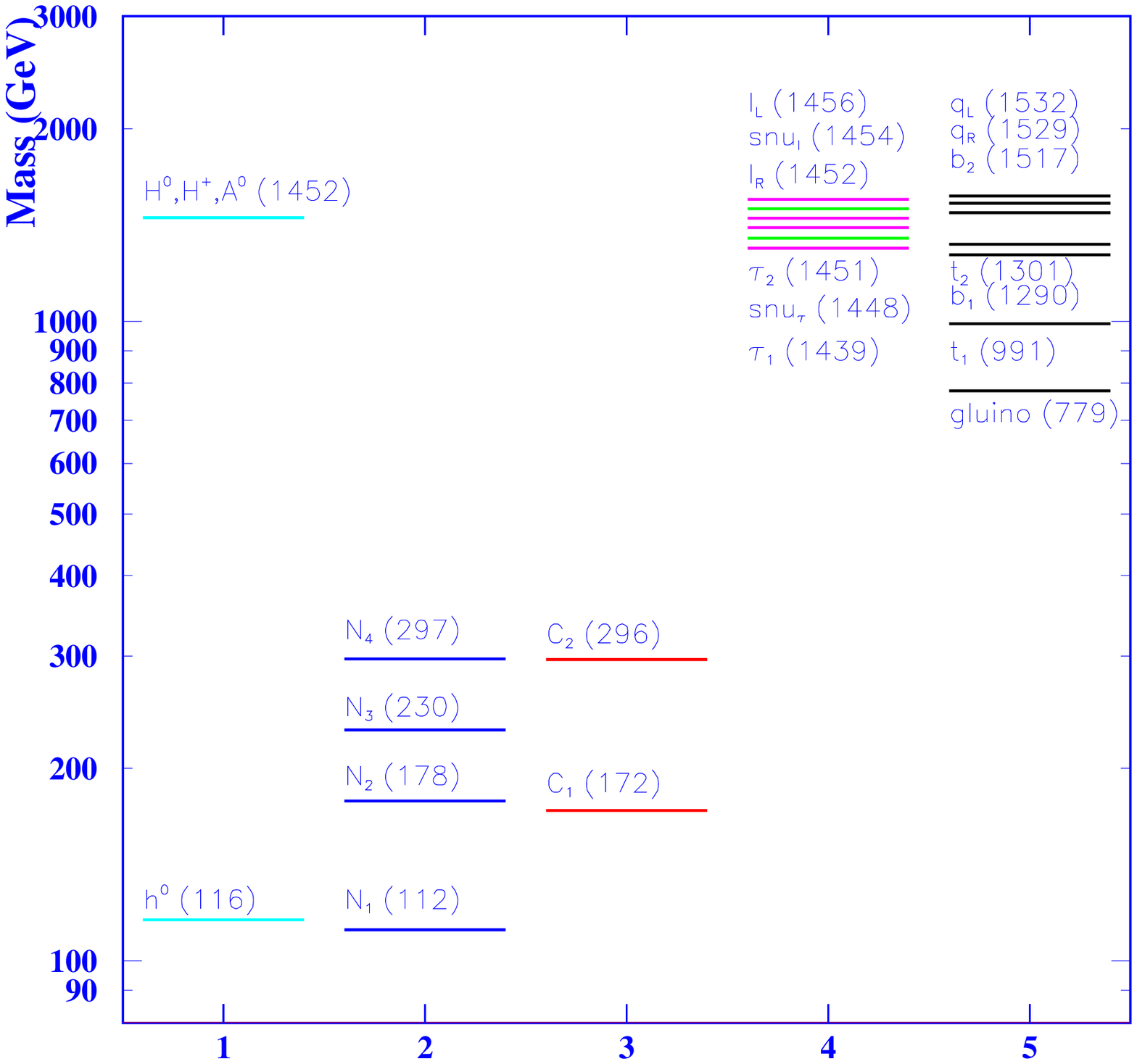,bbllx=0,bblly=10,bburx=520,bbury=500,width=8.0cm,clip=}
\caption{(Left) Qualitative overview of the locations the proposed
CMSSM benchmark points in a generic ($m_{1/2},m_0$) plane.
(Right) Example of a mass spectrum for point E of \protect 
\cite{ellis}.}
\label{fig:crossx}
\end{center}
\end{figure}

\begin{figure}
\begin{center}
\epsfig{file=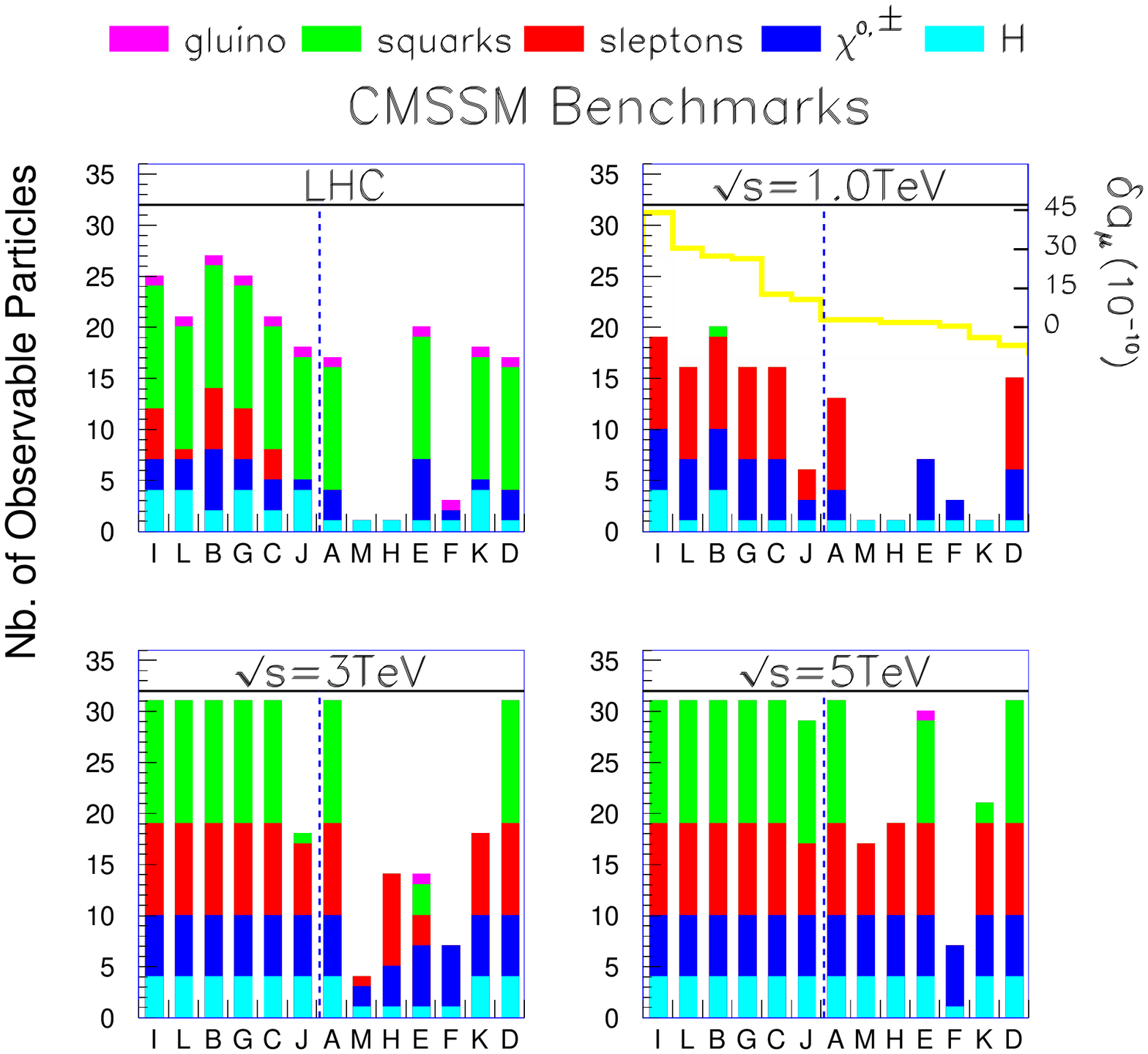,bbllx=0,bblly=0,bburx=580,bbury=580,width=8cm,clip}
\epsfig{file=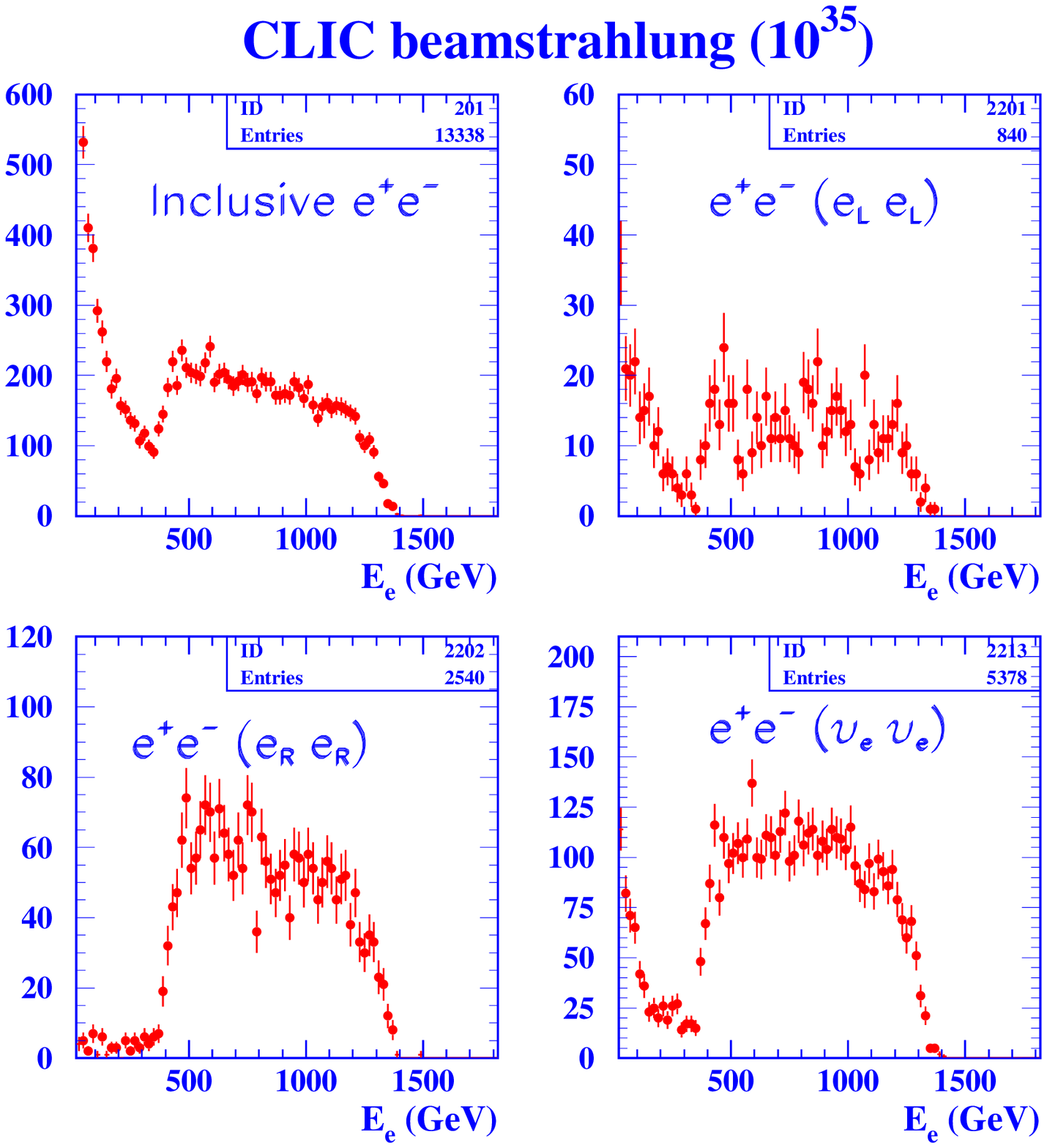,bbllx=0,bblly=150,bburx=580,bbury=750,width=8cm,clip}
\caption{(Left) 
Comparison of the capabilities of various colliders to
observe
different species of supersymmetric particles, in a range of
supersymmetric models whose parameters are chosen to be compatible with  
experimental constraints from LEP and elsewhere.  The models
are
ordered by their degree of compatibility with the recent BNL measurement
of the muon anomalous magnetic moment, as indicated by the line in the top
right panel. We see that linear colliders complement the LHC, via their
abilities to observe weakly-interacting sparticles, in particular. In most
of the restricted set of simplified models studied, CLIC (almost)
completes the supersymmetric spectroscopy initiated by the LHC.
(Right) Measured electron and positron spectrum for inclusive
 di-electron events for point E, for $\sqrt{s} = 3.5$ TeV 
and 650 fb$^{-1}$ using SPYHTIA, CALYPSO and SIMDET. Panels 2, 3 
and 4 show the di-electron events from $\tilde{e}_L\tilde{e}_L,
\tilde{e}_R\tilde{e}_R$ and $\tilde{\nu}_e\tilde{\nu}_e$ assuming ideal process
separation~\protect 
\cite{ellis,wilson}.
}
\label{fig:susy2}
\end{center}
\end{figure}

Studies for future facilities rely on benchmark points.
The  benchmark points used 
in previous studies
 have by now been mostly ruled
out  by the 
final measurements at LEP and the Tevatron. 
Recently a new set of benchmark points has been suggested
in~\cite{ellis}.
This study is based  on the constrained MSSM model
and takes  into account
the experimental search limits on sparticles, Higgs bosons, $\tan\beta$,
 $b \rightarrow s\gamma$ results,
 the compatibility with Dark Matter $(0.1 <\Omega h^2< 0.3)$,
and, for some points, the compatibility with $g-2$.
The model parameters
$m_{1/2}, m_0, \tan\beta$, sign of $\mu$
have been varied, while in order to not inflate the parameter space
too much
only points with $A=0$ have been chosen.
An overview of the selected points in the 
 $m_0, m_{1/2}$ plane is given in 
Fig.~\ref{fig:crossx}. The light coloured regions in which the 
points are shown,
 are the allowed regions; these include regions towards large $m_{1/2}$ with
fast coanihilation regions, a
rapid annihilation funnel,  and a focus point region close to the 
border of the region 
where electroweak symmetry breaking is no longer possible.  
Note that the points are chosen to span a large part of the 
phase space, including the more extreme corners, and cover $\tan \beta$ values
from 5 to 50. 
Some of the points are 3$\sigma$ away from the
value of $g-2$ that was reported in early 2001, 
 and have large fine 
tuning values. Hence, these  points do  
 {\it not} represent  a sampling of the region according to the `likeliness'
or compatibility with constraints, but rather represent a collection of 
 different scenarios, and are  thus not 
meant for a statistical study.

All of the scenarios proposed  in~\cite{ellis} are found to have sparticles
with masses larger than 500 GeV.   Thus, a LC with an energy 
above 1 TeV would be needed to study all sparticles in $e^+e^-$ collisions.
An example of the mass spectrum is shown in Fig.~\ref{fig:crossx}
 for 
point (E) of ~\cite{ellis}, 
corresponding to 
$m_{1/2}= 300 $ GeV, $m_0=1450 $ GeV, $\tan\beta=10$, sign of $\mu>0$
and $A=0$. For this point the heavy Higgses, sleptons and squarks 
have masses lager than 700 GeV.

We do not yet know which, if any, SUSY scenario is realized in Nature 
and where we should search for the SUSY particles. The LHC will certainly
tell us if SUSY particles exist at the TeV scale,
but the chances are large that at least
some of the sparticles will be out of reach at a 1 TeV collider. 
Sooner or later the need will arise to measure as precisely
as possible {\it all} members of the sparticle spectrum, in order to 
disentangle the underlying theory and (SUSY breaking) dynamics as completely 
as possible. It has been shown for TeV class LC SUSY sparticle measurements,
that knowing the masses to  
a level better than 1\% rather than about 10\%  (typical
for the LHC) is an enormous help for bottom-up approaches to 
reconstruct the SUSY breaking mechanism~\cite{blair}.
Another important aspect of $e^+e^-$ linear colliders is initial state
beam polarization, which can be used to single out newly produced states
of definite helicity, and further disentangle SUSY 
parameters~\cite{moortgat}.


Since supersymmetry will have been discovered (or 
possibly refuted) already 
before the turn on of a multi-TeV collider, its main task 
could be to complete the sparticle spectrum with precise measurements.
The number of particles which can be detected at a LC and LHC
for the chosen SUSY benchmark points is shown in Fig.~\ref{fig:susy2}.
For a sparticle to be observed at a LC, its cross section has to be larger than
0.1 fb$^{-1}$.
The points are ordered according to decreasing value of their compatibility with
$g-2$. For the `$g-2$ friendly' points the LHC can detect the squarks, 
and a 1 TeV collider can often
detect  sleptons and gauginos, and sometimes also 
the heavy Higgs particles. 
The notion of `$g-2$ friendly' is, however, somewhat changing with time. 

A multi-TeV collider can almost always produce 
all sparticles, except the gluino, for $g-2$ friendly points, and also 
quite a few sparticles from points in the
$g-2$ `unfriendly' region. It is clear that with our present lack of 
knowledge of the mass pattern of the sparticles, the option to 
upgrade a multi-TeV collider like CLIC to 5 TeV 
must be kept. Moreover, Fig.~\ref{fig:susy2} shows that even  a 5 TeV
collider
will not cover the full phase space of presently allowed scenarios.
In order to cover all points proposed in~\cite{ellis} --which does
not necessarily give the upper limit of the possible sparticle
mass range either--
a $e^+e^-$ collider of 8 TeV would be necessary, and the 
luminosity would need to be increased by a factor 2-5 to produce
measurable samples of all sparticles.

Next one needs to study the question whether a machine like CLIC, with its
important backgrounds and luminosity spectrum smearing can make 
precision measurements of sparticle properties.
A case study is shown in Fig.~\ref{fig:susy2} for sneutrino pair production
and the sneutrino mass determination for point (E) above.
The signal is 
$\tilde{\nu}_e\tilde{\nu}_e \rightarrow 
e^+\tilde{\chi}_1^- e^-\tilde{\chi}_1^+$~\cite{wilson}.
The study relies on the typical `box' shape of the inclusive lepton spectrum, 
a technique developed for  a TeV class LC: the end points of the rectangular
shape  in Fig.~\ref{fig:susy2}
determine the slepton and gaugino mass. 
The study includes background processes and machine background effects.
The figure shows that
 the signal is preserved in the CLIC environment.
A recent study ~\cite{battpriv} shows that 
 mass measurements for smuons are possible with a precision of $O(1\%)$.

In general the gluino is difficult to  measure at an 
$e^+e^-$ LC, except when it is lighter than the squarks.
A possible way to measure  gluinos is via the $\gamma\gamma$ collider 
option. In~\cite{klasen} it was shown that gluinos are produced with 
cross sections which are a few orders of magnitude larger than in $e^+e^-$
collisions.
Gluinos could be detected at a 3 TeV collider in the $\gamma\gamma$ mode 
for masses up to about 1 TeV, where they 
are produced  with cross sections in the range
of 1-0.1 fb$^{-1}$,
if the gluinos are heavier than the squarks.
Hence  the initial studies indicate that if SUSY is stabilizing the
SM, and sparticles have masses in the anticipated range, then a multi-TeV
$e^+e^-$ collider would be the ideal tool to complete the 
sparticle spectrum.
 
%

\begin{figure}
\begin{center}
\epsfig{file=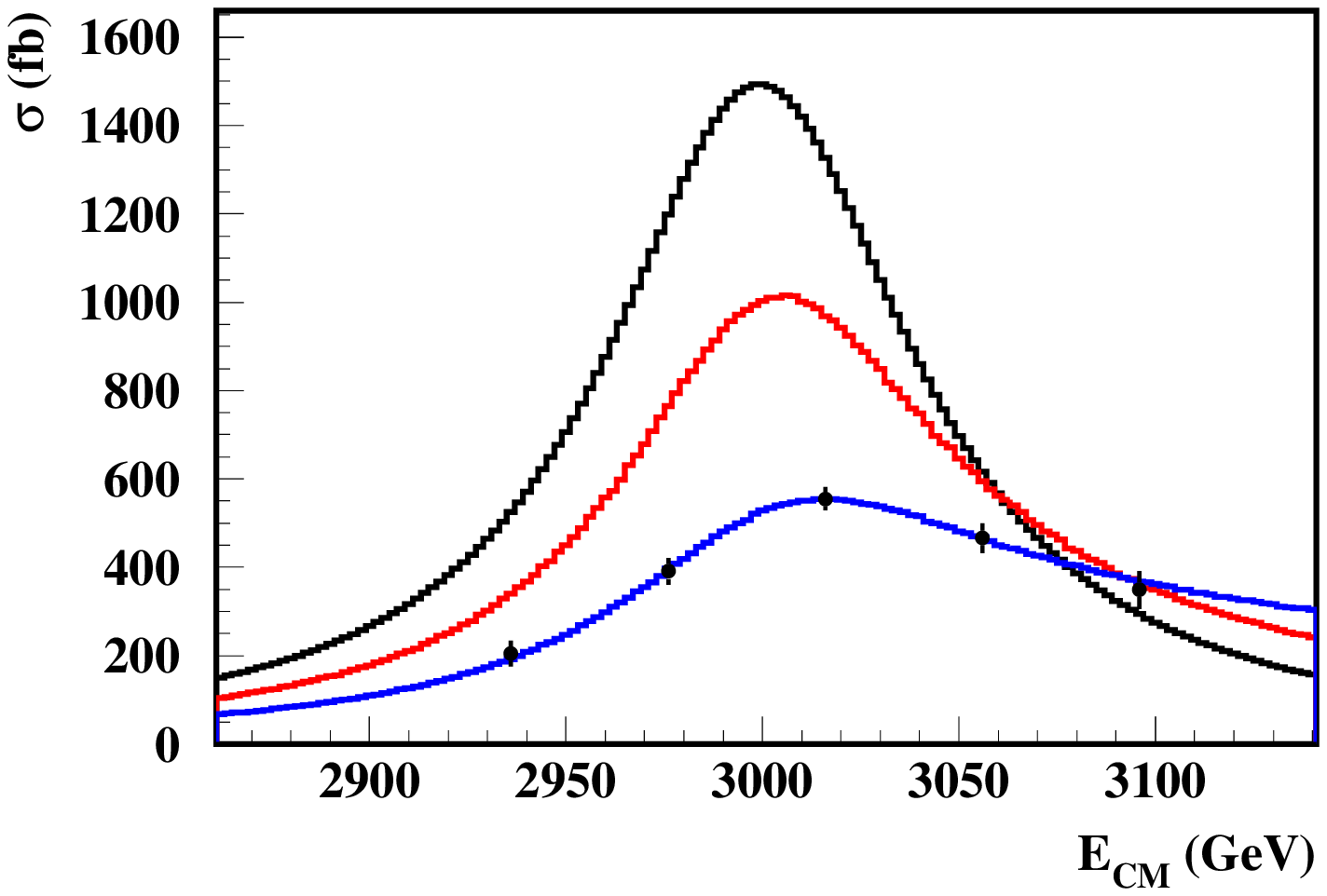,bbllx=15,bblly=35,bburx=430,bbury=330,width=8cm,height=7cm,clip}
\epsfig{file=e3_002_bess_afb_016.eps,bbllx=0,bblly=0,bburx=490,bbury=550,width=8cm,clip}
\caption{
(Left) Production of $Z^{'}\rightarrow l^+l^-$ 
shown at the Born level (top), 
including ISR (middle)  and further including 
smearing due to  the luminosity spectrum.
(Right) Hadronic cross section and Forward-Backward asymmetries at energies 
around 3 TeV. Continuous lines are predictions by the D-BESS model with 
$M$ = 3 TeV and $g/g^{'' }= 0.015$, and dots the observable 
D-BESS signal after including smearing of the  luminosity 
spectrum~\protect \cite{battagz}.}
\label{fig:zprime}
\end{center}
\end{figure}

\section{$\ee\ra f\bar f$ and Precision Measurements}


Extending the sensitivity of new physics beyond the reach of the 
LHC is a prime aim for a future collider.
LEP has demonstrated how an $e^+e^-$ collider can provide
 precision measurements, and be a powerful tool for searching for 
new physics. Reactions at $\ee$ colliders are simple, clean,
and precisely calculable.  These features enable an $\ee$ collider
to match, and in many instances surpass, the energy scale reach
of contemporaneous higher energy hadron colliders.

There is a wide range of new physics scenarios predicting the existence of new
 vector 
particles with masses in the TeV range.
These basically  
aim at explaining the origin of electroweak symmetry breaking if there 
is no light 
elementary Higgs boson, stabilizing the SM if SUSY is not realised in 
Nature, or 
embedding the SM in a theory of grand unification.

One of the simplest extensions of the SM involves the introduction
of  an additional 
$U(1)$ gauge symmetry, whose breaking scale is close to the Fermi scale,
leading to new gauge bosons.
Several  models for  the production of new $Z'$ bosons have been 
studied for a multi-TeV collider~\cite{battagz}.
First we study the case of direct production of such new bosons, 
which will produce a resonance signal in the two-fermion production
processes, just like the $Z$ at LEP. 
The LHC could also produce such bosons with a detectable rate 
 in the range up to a few 
TeV, but it will not have the statistical power to measure in detail 
their properties.

For the multi-TeV collider study we assume
 a new gauge boson with 
 $M_{Z^{'}}$ of 3.0 TeV and a width $\Gamma(Z^{'})/M_{Z^{'}} \simeq
\Gamma(Z^{0})/M_{Z^{0}}$. A  `LEP-like' scan
 of the cross section in five points, as shown in
Fig.~\ref{fig:zprime}, for a total of 1 ab$^{-1}$
at CLIC gives the following 
fit accuracy:  $\delta M_Z'/M_Z' \sim 10^{-4}$ and
$\delta \Gamma_Z'/\Gamma_Z' \sim 3 \cdot 10^{-3}$ .
The precision is somewhat worse than the one obtained for the $Z$ 
at LEP, but would still be a textbook 
measurement.

Some of these models, such as the   degenerate BESS Model, predict
several almost degenerate resonances in the multi-TeV region. 
Disentangling these resonances and measuring their properties will be
particularly challenging.
For example,  in the degenerate BESS model one expects
 two almost degenerate triplets: $L_3, L_3^{\pm}$ and $R_3, R_3^{\pm}$.
The sensitivity to $L_3$ and $R_3$ with $M=$3~TeV
for $\cal{L}=$500~fb$^{-1}$  at LHC and
$\cal{L}=$1~ab$^{-1}$ at CLIC are given in Table~\ref{tab:zprim}.
A possible 
 energy scan of narrow resonances is shown in 
Fig.~\ref{fig:zprime} for the case of  $g/g''=0.15$, with 
$g''$ the coupling corresponding to the additional
symmetry. The studies show that
 CLIC can distinguish resonances 
with a  $\Delta M$ down to 13 GeV, corresponding to $g/g''>0.08$
in this model.

\begin{table}
\begin{center}
\caption{Sensitivity to $L_3$ and $R_3$ production at the 
LHC (${\cal{L}}=500 $ fb$^{-1}$)
and CLIC (${\cal{L}}=1 $ ab$^{-1}$).}
\begin{tabular}{|c|c|c|c|c|c|c|}
\hline
$g/g''$ & $M$ & $\Gamma_{L_3}$ / $\Gamma_{R_3}$ & $S/\sqrt{S+B}$&
 $S/\sqrt{S+B}$ & $\Delta M$
\\
& (GeV) &(GeV)  & LHC ($e+\mu$) & CLIC (had.) &  CLIC \\
 \hline 
0.1 & 3000 & 2.0 / 0.3 & 3.4& ~62 & 23.20 $\pm$ .06
\\
0.2 & 3000 & 8.2 / 1.2 & 6.6& 152 & 83.50 $\pm$ .02
\\
\hline
\end{tabular}
\label{tab:zprim}
\end{center}
\end{table}

Also, the direct production of new gauge bosons 
in the Left-Right symmetric model has been studied!\cite{ferrari2}; 
in particular 
the production of a 
$Z'$ with a mass of 3 TeV
which decays into heavy Majorana neutrinos.
This leads to multi-jet plus lepton final states. The large number of 
$Z'\rightarrow N_eN_e$ events can be used to determine the masses of these 
new particles with an accuracy of 0.01\% for the $Z'$ and 0.19\%
for the heavy neutrino.

Apart from the direct observation of new resonances, signals of 
this kind of new physics at even higher energy scales  can be 
discovered in precision measurements of two fermion production
processes. The sensitivity to new physics  
of
LEP observables such as 
$\sigma_{f \bar f}$, $A_{FB}^{f \bar f}$ and $A_{LR}^{f \bar f}$
has been studied with the CLIC tools.
The statistical accuracy for the determination of 
$\sigma_{f \bar f}$, $A_{FB}^{f \bar f}$ and $A_{LR}^{f \bar f}$ has been 
determined, for $\mu^+\mu^-$ and $b \bar b$, taking the CLIC parameters at
$\sqrt{s}$ = 3~TeV.
The relative statistical accuracies for these observables,
$\delta {\cal{O}}/{\cal{O}}$, are given in Table~\ref{tab:res}.
Note that these estimations take full account of the background at CLIC.

\begin{table}
\caption{Relative statistical accuracies on electroweak observables, obtained for 
1~ab$^{-1}$ of CLIC data at $\sqrt{s}$ = 3~TeV, including the effect of 
$\gamma \gamma \rightarrow {\mathrm{hadrons}}$ background.}  
\begin{tabular}{|l|c|}
\hline
Observable & Relative Stat. Accuracy \\
           & $\delta {\cal{O}}/{\cal{O}}$ for 1~ab$^{-1}$ \\
\hline \hline
$\sigma_{\mu^+\mu^-}$ & $\pm 0.010$ \\
$\sigma_{b \bar b}$ & $\pm 0.012$ \\
$A_{FB}^{\mu\mu}$ & $\pm 0.018$ \\
$A_{FB}^{bb}$ & $\pm 0.055$ \\ \hline
\end{tabular}
\label{tab:res}
\end{table}

\begin{figure}
\begin{center}
\begin{tabular}{l c r}
\includegraphics[width=0.44\textwidth,height=0.4\textwidth,bbllx=0,bblly=0,bburx=580,bbury=550]{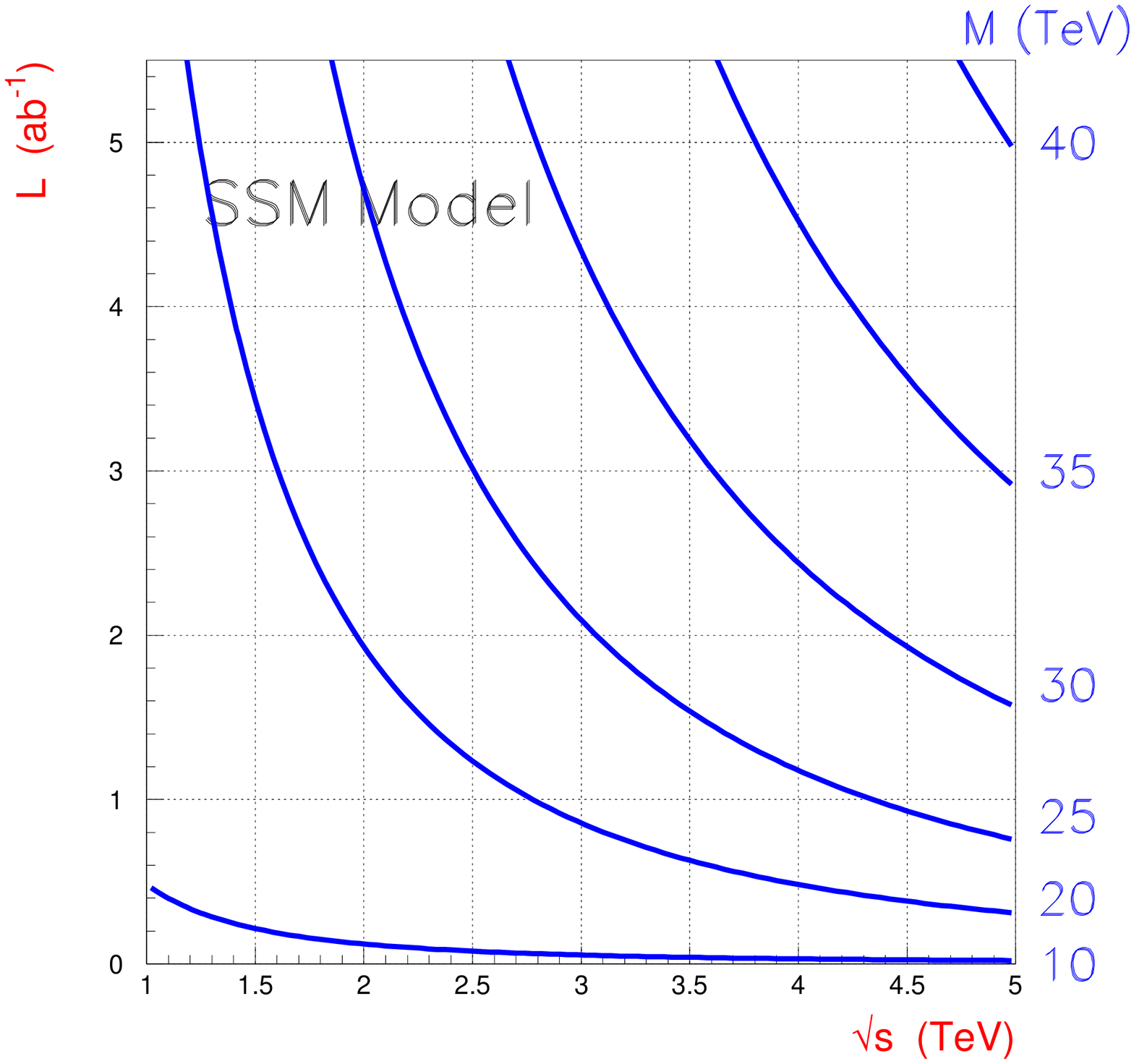}&
\includegraphics[width=0.44\textwidth,height=0.4\textwidth,bbllx=0,bblly=0,bburx=580,bbury=550]{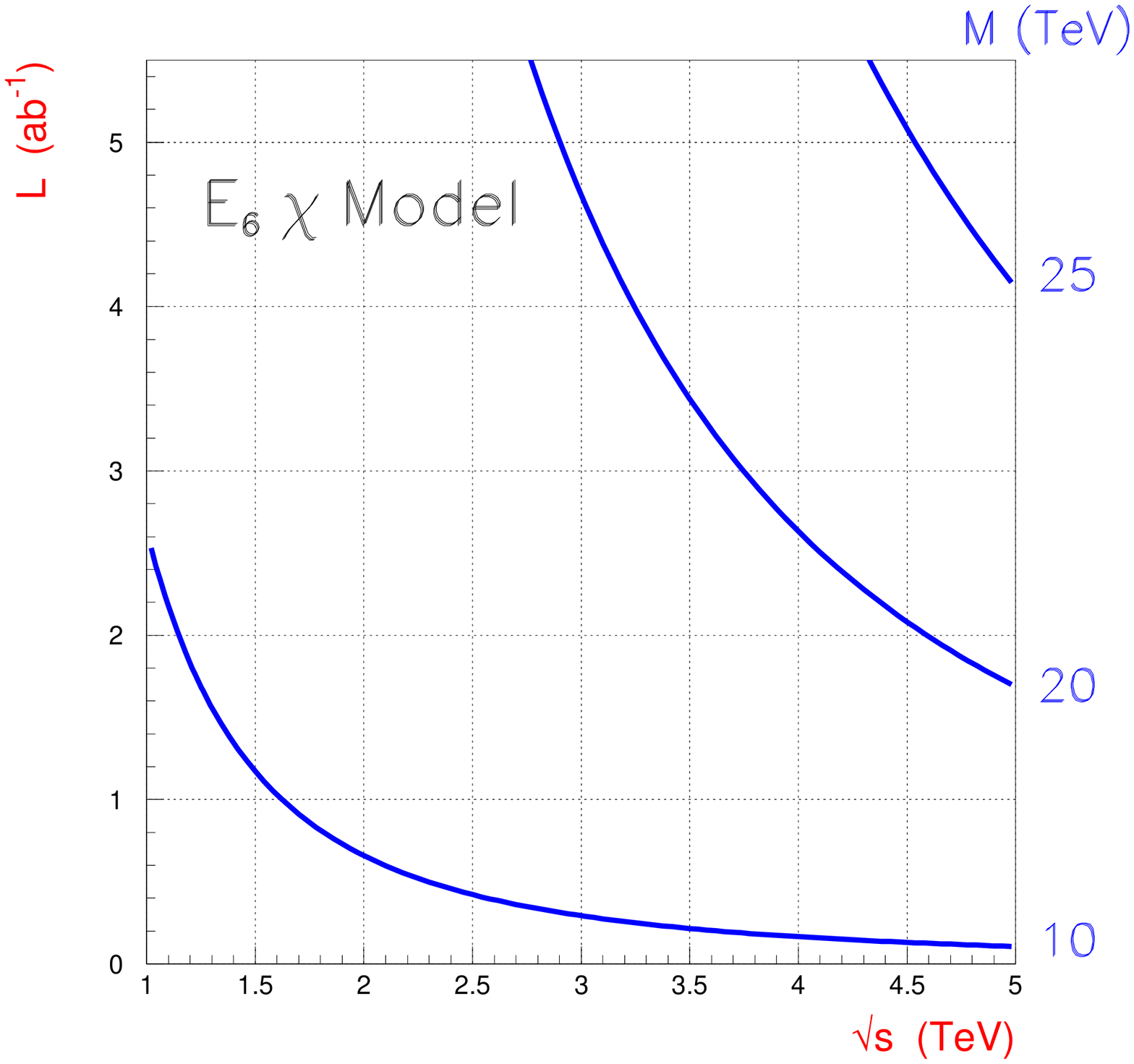}&
\end{tabular}
\end{center}
\caption{The 95\% C.L. sensitivity contours in the $\cal{L}$ vs. 
$\sqrt{s}$ plane
for different values of $M_{Z'}$ in the SSM model (left)
and in the $E_6$~$\chi$ 
model (right)~\protect \cite{battagz}.}
\label{fig:zp}
\end{figure}

The sensitivity to search for indirect signals of $Z'$ 
bosons increases
with luminosity and 
available CMS energy:
\begin{equation}
M_{Z'} \propto \sqrt{s \frac{
{\sigma}}{\delta {\sigma}}}
\propto \sqrt{s \sqrt{\frac{{\cal{L}}}{s}}} = (s \times {\cal{L}})^{1/4}
\label{resc}.
\end{equation}
Hence a 5 TeV collider with ${\cal L} \cong 
10^{35}$cm$^{-2}$s$^{-1}$ can achieve a factor 3 to 4 more in 
discovery reach 
compared to  a TeV
class collider.
These scaling laws have been verified by detailed studies using the 
accuracies of the two-fermion related measurements as given in 
Table~\ref{tab:res}. The results based on the detailed 
studies,  for two different models which 
produce a $Z'$, are shown in Fig.~\ref{fig:zp}.
For the SSM model, a region up to $M_{Z'} = 40 $ TeV can be 
covered for 5 ab$^{-1}$~\cite{battagz}.

\begin{figure}[htbp]
  \begin{tabular}{lr}
   \includegraphics[width=0.5\textwidth,height=0.55\textwidth,bbllx=0,bblly=0,bburx=430,bbury=570]{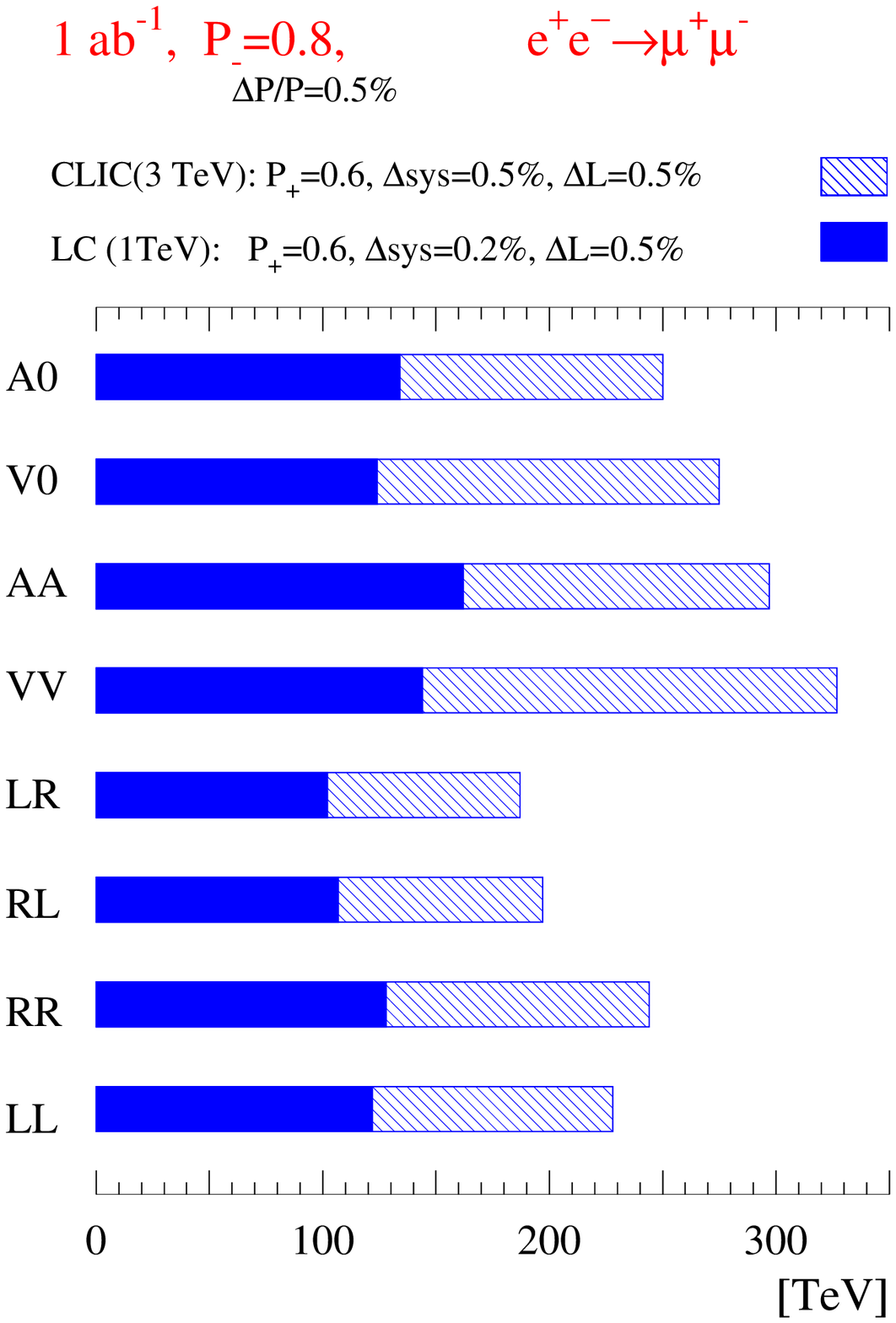}&
   \includegraphics[width=0.5\textwidth,height=0.55\textwidth,bbllx=0,bblly=0,bburx=430,bbury=570]{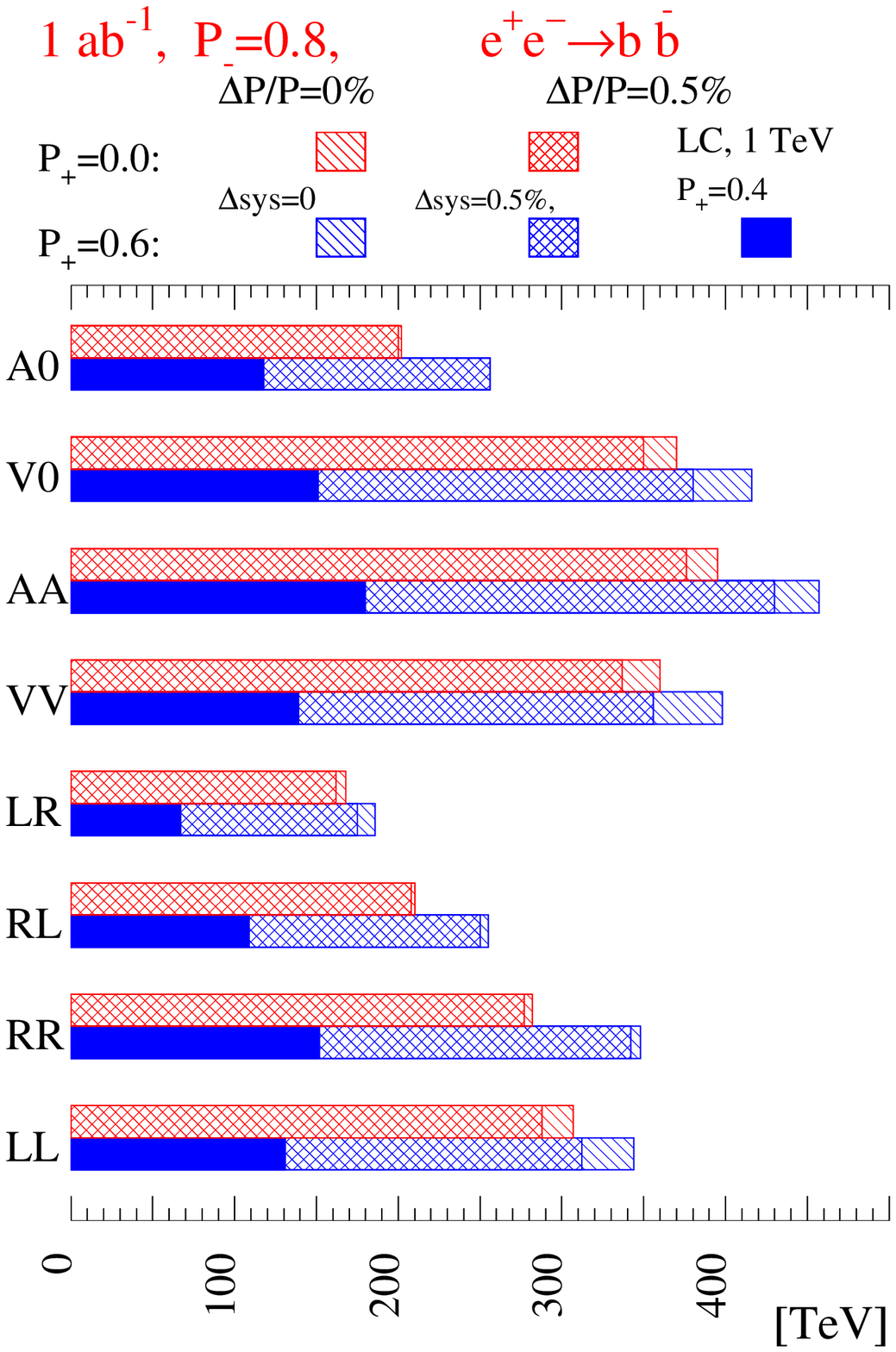}
  \end{tabular}
\caption{Limits on the scale $\Lambda$ of contact interactions in 
$ee\rightarrow \mu\mu,b \bar b$
for CLIC operating at 3~TeV (dashed histogram) compared to a 1~TeV LC 
(filled histogram) for different models and the 
$\mu^+\mu^-$ (left) and $b \bar b$ 
(right) channels. The electron 
polarizations ${\cal{P}}_{-}$ is taken to be 0.8 and 
the positron ${\cal{P}}_{+}$ to be 0.6. 
For comparison the upper bars in the right plot
show the sensitivity achieved without positron 
polarization~\protect \cite{battagz}.}
\label{fig:ci}
\end{figure}

\begin{figure}
\begin{center}
\epsfig{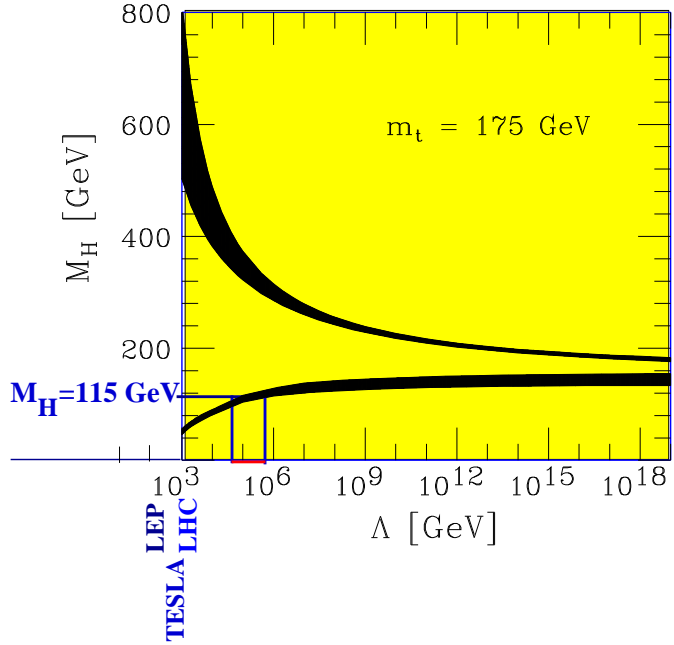}
\end{center}
\caption{The range allowed for the mass of the Higgs
boson if the SM is to remain valid up to a given scale $\Lambda$.
In the upper part of the plane, the effective potential  blows up,
whereas in the lower part the present electroweak vacuum is 
unstable~\protect \cite{stabeliz}.}
\label{fig:lighthiggs}
\end{figure}

The studies above have addressed specific models of new physics 
beyond the SM.
Fermion compositeness or exchange of very heavy new particles can be 
described more  generally by four-fermion contact interactions \cite{ref:ci}. 
These parameterize the
interactions beyond the SM by means of an effective scale, $\Lambda$,
\begin{equation}
{\cal L}_{CI} =             \sum_{i,j = \mathrm{L,R}} \eta_{ij}
           \frac{g^2}{\Lambda^2_{ij}}
           (\bar{\mathrm{e}}_i \gamma^{\mu}\mathrm{e}_i)
           (\bar{\mathit{f}}_j \gamma^{\mu}\mathit{f}_j).
\label{ci_lagr}
\end{equation}
By convention one usually takes $g^2/4\pi=1$, and scenarios with 
different chirality are distinguished from one another 
by specifying  values of $\pm 1$ or 0 for the 
coefficients  $\eta_{ij}$. 
The contact scale $\Lambda$ can then be interpreted 
as the mass of the new particles exchanged by the strongly interacting
fermion constituents.

%

The sensitivity 
of electroweak observables to the contact 
interaction scale $\Lambda$ was estimated using the 
statistical accuracies in Table~\ref{tab:res}  
for the $\mu \mu$ and $b \bar{b}$ final states. 
Beam polarization represents an important tool in these studies. 
First, it improves 
the sensitivity to new interactions through the introduction of 
the left-right 
asymmetries $A_{LR}$ and the polarized forward-backward asymmetries 
$A_{FB}^{pol}$ in the electroweak fits. If both beams can be polarized to 
${\cal{P}}_{-}$ 
and ${\cal{P}}_{+}$ respectively, the relevant parameter is the 
effective polarization defined as 
${\cal{P}} = (-{\cal{P}}_{-}+{\cal{P}}_{+})/
({1-{\cal{P}}_{-}+{\cal{P}}_{+}})$.
In addition to the improved 
sensitivity, the uncertainty on the effective polarization,
can be made smaller than 
the error on the individual beam polarization measurements.
Secondly, in the case that 
 a significant deviation from the SM prediction is
observed, $e^-$ and $e^+$ 
polarization would be  useful in determining the 
helicity structure  of
the new interactions. 

Results are given in Fig.~\ref{fig:ci}
in terms of  lower limits on $\Lambda$ which can be excluded 
at 95\% C.L. 
High luminosity $e^+e^-$ collisions at 3~TeV can 
probe $\Lambda$  scales in the range 200-400~TeV for 1 ab$^{-1}$.  
For comparison, 
the corresponding results expected for a LC operating at 1~TeV are also 
shown~\cite{battagz}.
Using the scaling law, the expected gain in reach on $\Lambda$ for
 5 ab$^{-1}$ and
a 5 TeV (10 TeV) $e^+e^-$ collider would be 400-800 GeV (500-1000) TeV.
This is a very exciting prospect, {\it if} the `doomsday'
scenario is encountered, where some years from now  only a light Higgs has been 
discovered, and no sign of other new physics has been revealed by the 
LHC or a TeV class LC.
Indeed, if the Higgs particle 
 is light, i.e. below 150 GeV or so, then the Standard Model
cannot be stable up to the GUT or Planck scale, and a new mechanism
 is needed to stabilize it as shown in
Fig.~\ref{fig:lighthiggs}~\cite{stabeliz}: only a narrow corridor up to the 
Planck scale is allowed for Higgs masses around 180 GeV.
For example, for a Higgs with a mass in the region of 115-120 GeV, the 
SM will hit a region of electroweak unstable vacuum in the range of 
100-1000 TeV.
Hence, if the theory assessment of Fig.~\ref{fig:lighthiggs} remains
valid, and 
the bounds do not change significantly (which could happen due to a change
in the top-mass from e.g. new measurements at the Tevatron) {\it and} 
the  Higgs is as light as  120 GeV, then the signature of new physics 
cannot escape precision measurements at a multi-TeV collider.

Another example of precision measurements is the  determination of 
triple gauge boson couplings, studied e.g. via $e^+e^- \rightarrow
W^+W^-$. Deviations from the SM expectations in  these couplings 
are a sign of new physics. An initial study, using TeV class LC results for 
background and detector effects, suggests 
that the sensitivity
 of the
anomalous $\Delta \lambda_{\gamma}$ and $\Delta \kappa_{\gamma}$ couplings
(which are zero in the SM) for CLIC at 3 (5) TeV amounts to 
roughly $1.3 \cdot 10^{-4} (0.8 \cdot 10^{-4})$ 
and $0.9 \cdot 10^{-4} (0.5 \cdot 10^{-4})$ respectively, for 1 ab$^{-1}$
of data~\cite{barklowtgc}.
Hence a multi-TeV collider can probe these couplings a factor 2-4 times more 
precisely than a TeV class LC.


\begin{figure}[htbp]
\epsfig{file=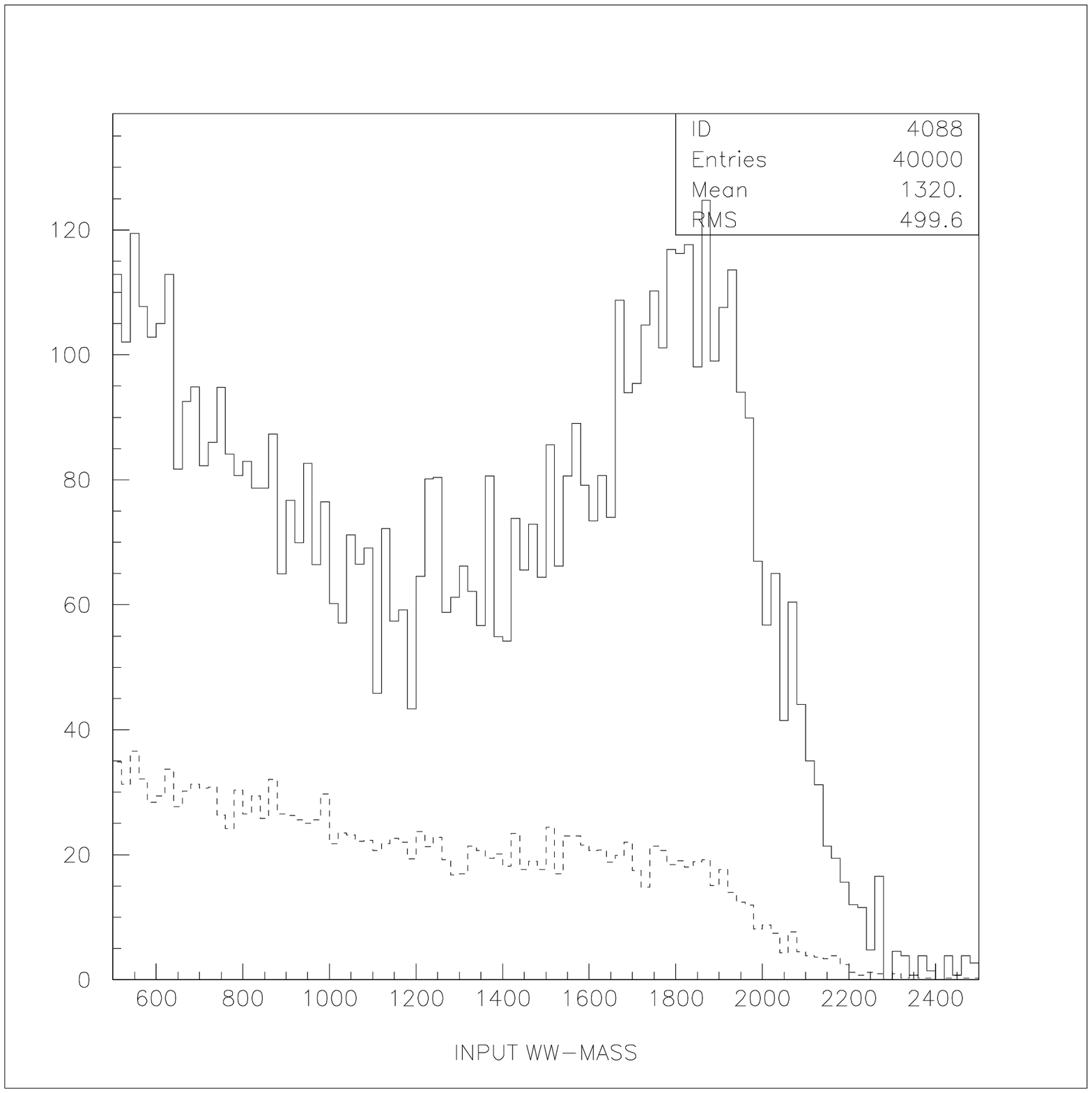,bbllx=30,bblly=150,bburx=560,bbury=680,width=7cm,clip=}
\epsfig{file=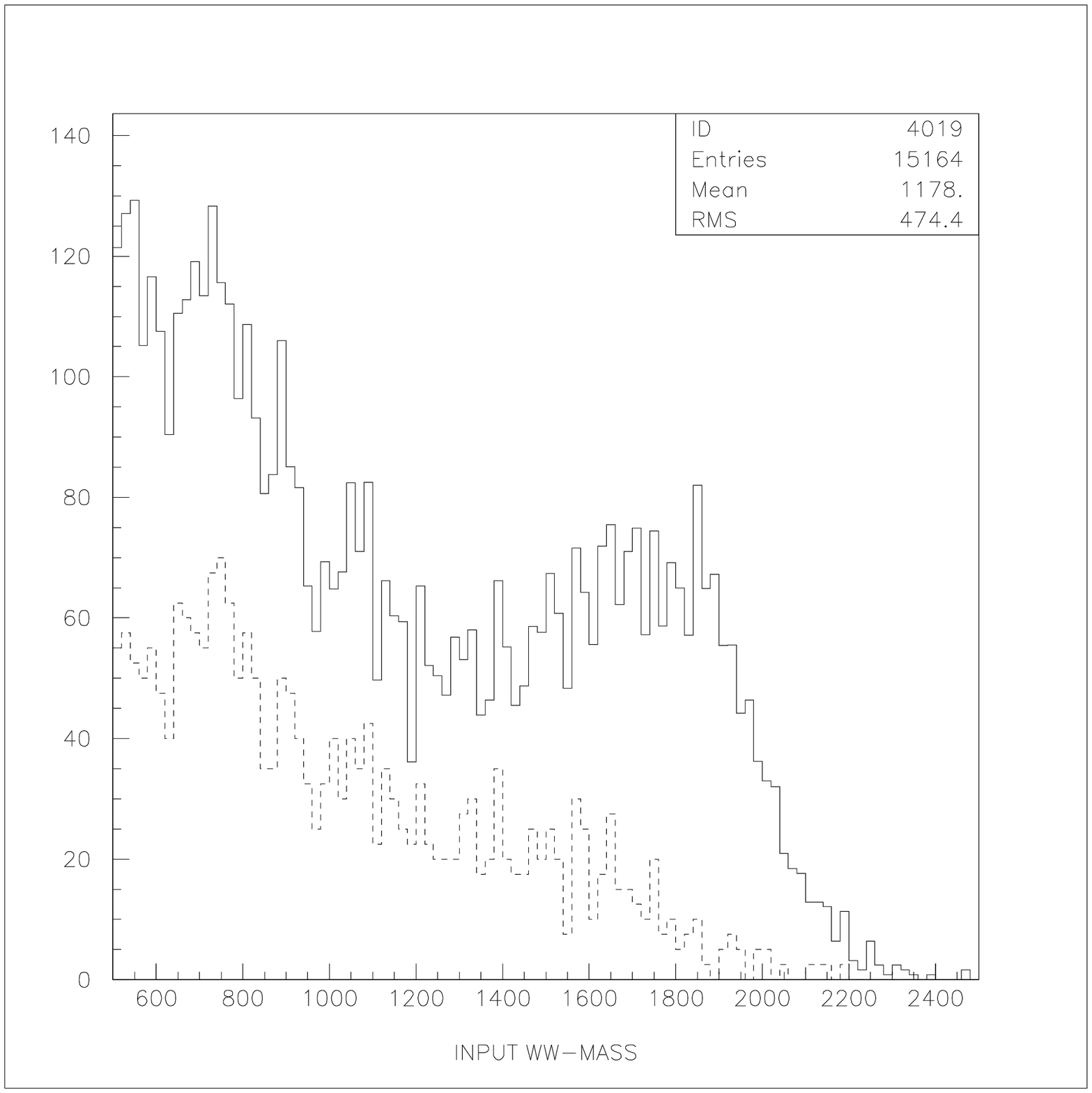,bbllx=30,bblly=150,bburx=560,bbury=680,width=7cm,clip=}
\caption{Mass spectrum for $WW$ scattering and $WZ$ and $ZZ$ scattering 
backgrounds for $\alpha_5=-0.002, \alpha_0 = 0.0$, before detector (left)
and after detector smearing and addition of $\gamma\gamma$ background
(right)~\protect \cite{deroeck}.}
\label{fig:wwfull}
\end{figure}

\begin{figure}[htb!] 
\includegraphics[height=7.0cm]{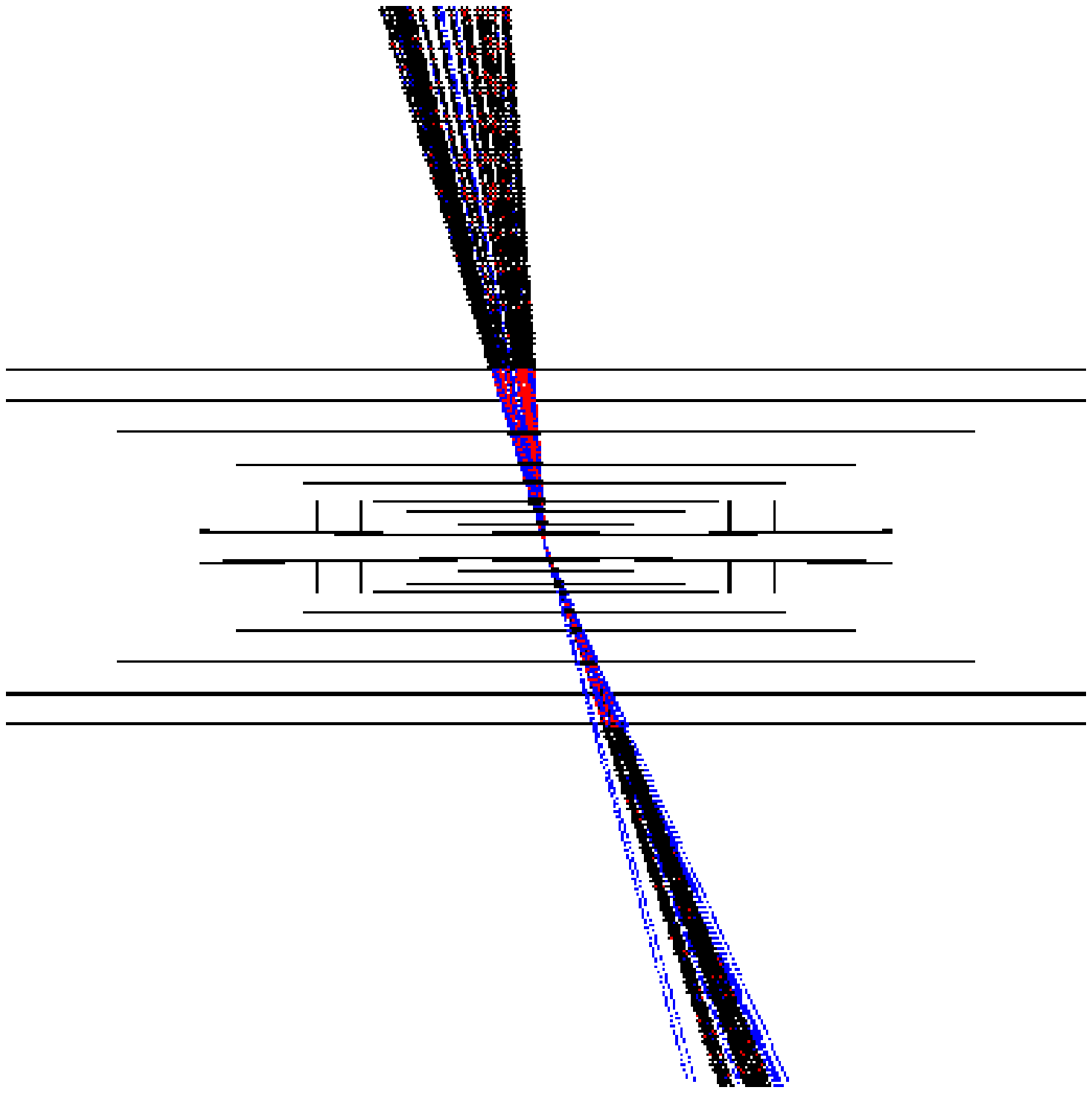}
\includegraphics[height=7.0cm]{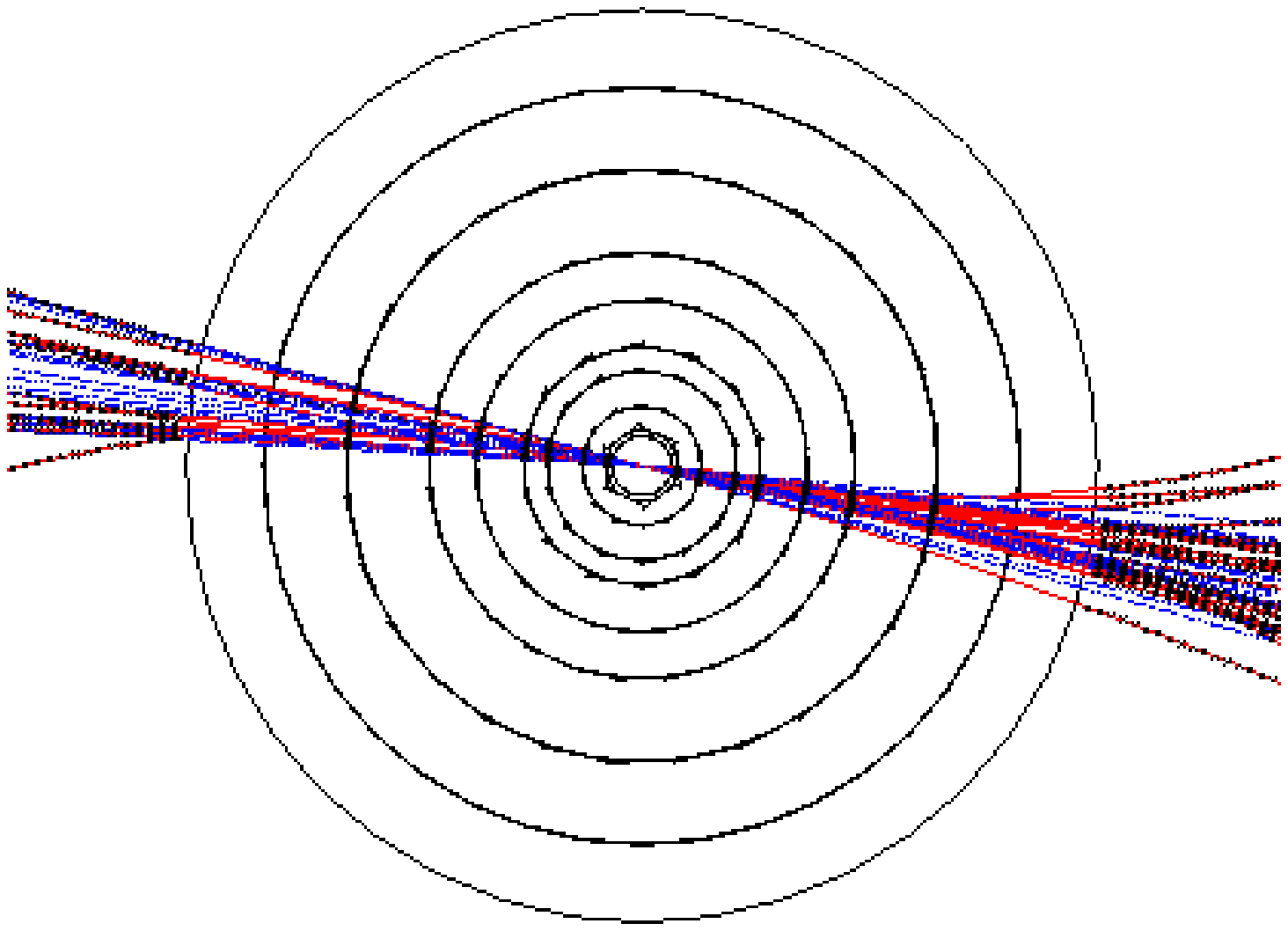}
\caption{Two views of an event in the central detector, of the type
$e^+e^-\rightarrow WW\nu\nu\rightarrow 4$ jets $\nu\nu$,  from a resonance
with $M_{WW}$ =2 TeV~\protect \cite{deroeck}.}
\label{fig:wwevent}
\end{figure}

\section{$\ww$ Scattering}


In the scenario that no
  Higgs boson with large gauge boson couplings  and a mass less than about 700 GeV is found, then the
$W^{\pm},Z$ bosons are expected to develop strong interactions
at scales of order 1-2 TeV. Generally one expects an excess of events
above Standard Model expectation, and, possibly, resonance formation.


The reaction $e^+e^- \rightarrow
\nu\overline{\nu}W^+_LW^-_L $ at 3 TeV was studied  using two approaches. 
The Chirally-Coupled Vector Model~\cite{barger} for $W_LW_L$
scattering describes the low-energy behaviour of a technicolor-type 
model with a Techni-$\rho$ vector resonance V(spin-1, isospin-1
vector resonance). 
The mass of the resonance can be chosen and the cases 
$M\sim 1.5, 2.0$ and 2.5 TeV were studied. 

In a second approach, the prescription given in ~\cite{bod,forshaw}
using the electroweak Chiral Langrangian (EHChL) Formalism is applied. 
The Higgs terms in the SM Langrangian are replaced by terms in the next order
of the Chiral expansion.
Unitarity corrections are important for energies larger than 1 TeV.
Here the Pade (Inverse Amplitude) protocol has been used.
The parameters $\alpha_4$ and $\alpha_5$ quantify our ignorance
of the new physics. 
High mass vector resonances will be produced in $WW$ and $WZ$ scattering 
for certain combinations of $\alpha_4$ and $\alpha_5$.


%


The total cross section for $WW\rightarrow WW$ scattering with the
values $\alpha_4 = 0.0 $ and $\alpha_5 =  -0.002$, amounts to 12
fb in $e^+e^-$ collisions at 3 TeV, and is measurable at a high
luminosity LC. With these parameters a broad resonance is produced
at 2~TeV in the $WW$ invariant mass, as shown in Fig.~\ref{fig:wwfull}.
A detector study is performed for this scenario, implemented
in the PYTHIA generator~\cite{forshaw}.
Events are selected with the following cuts:
$p^W_T > 150 $ GeV, $|\cos\Theta^W| < 0.8; M_{W} > 500$ GeV; $p_T^{WW} < 300$
GeV; and ($200 < M_{rec} < 1500 $ GeV), with $M_{rec}$ the recoil
mass.
Cross sections  including these  cuts for 
different masses and widths are calculated  with the program 
of~\cite{barger} and  are given in 
Table~\ref{tab:ww}.

The big advantage of an $e^+e^-$ collider is the clean
final state, which allows for the use 
of the hadronic decay modes of the $W$'s in selecting and reconstructing events.
Four jets are produced in the decay of the 
two $W's$. However, due to  the boost from the decay 
of the heavy resonance, the two jets of a $W$ are very
 collimated, and close to each other as shown
in Fig.~\ref{fig:wwevent}. 
With the present  assumptions on the energy flow in SIMDET, 
the resolution to reconstruct the $W,Z$
mass is about 7\%. 


A full spectrum which contains contributions from all the channels
$ZZ\rightarrow ZZ, ZZ\rightarrow WW,
ZW\rightarrow ZW, WW\rightarrow ZZ$ and $WW\rightarrow WW$ 
 is shown for 1.6 ab$^{-1}$ in Fig.~\ref{fig:wwfull}, before 
and after detector smearing with parameters 
$\alpha_5=-0.002, \alpha_0 = 0.0$~\cite{deroeck}.

Clearly
heavy resonances in $WW$ scattering can be detected at CLIC.
The signal is not heavily distorted by detector resolution and background.
Depending on the mass and width of the signal, about a 1000 events/year
could be fully reconstructed in the 4-jet mode at CLIC.
Good energy and track reconstruction will be important to remove 
backgrounds and reconstruct the resonance parameters (and hence the underlying
model parameters) accurately.

\begin{table}
\caption{ Cross sections for vector resonances in $WW$ scattering, 
with cuts as given in the text}
\begin{center}
\begin{tabular}{|c|c|c|c|}
\hline
$\sqrt{s}$ &  M=1.5 TeV & M=2.0 TeV & M=2.5 TeV\\
3 TeV & 
$\Gamma =$ 35 GeV & $\Gamma =$ 85 GeV & $\Gamma =$ 250 GeV \\
\hline
$\sigma$ (fb) & 4.5 & 4.3 &4.0 \\
\hline
\end{tabular}
\label{tab:ww}
\end{center}
\end{table}

\section{Extra Dimensions}

For the past  few years the phenomenology  
of large extra dimensions has been 
explored at the TeV scale. These theories aim to solve the hierarchy problem
by bringing the gravity scale closer to the  electroweak  scale.
(If you cannot get to the Planck scale, try to bring the Planck scale to you.)
The hidden extra dimensions can dramatically change
the strength of gravity, and bring the Planck scale down
to  the  TeV region.

\begin{figure}[htbp]
\centerline{
\includegraphics[width=5.4cm,angle=90,bbllx=80,bblly=80,bburx=530,bbury=730]{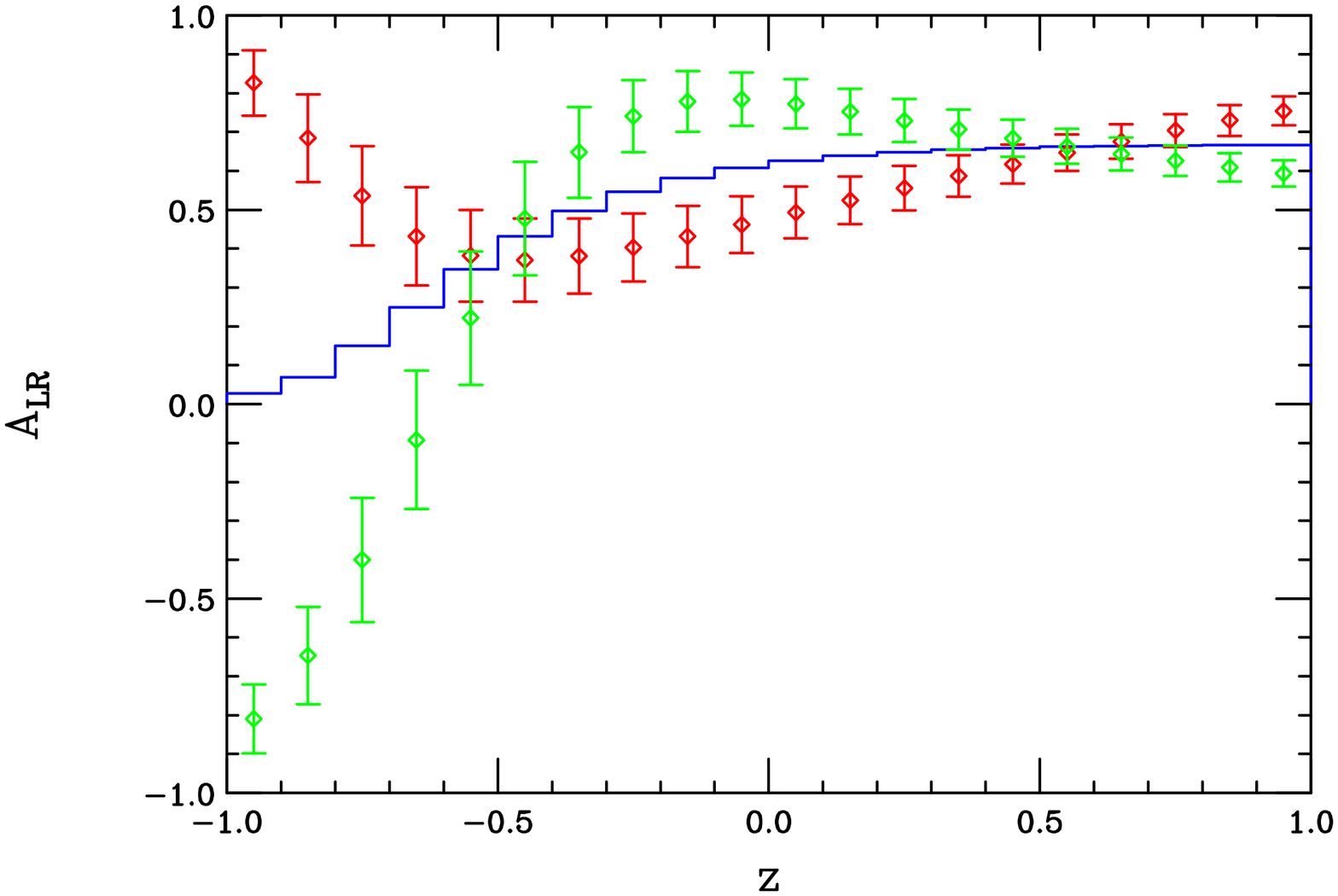}
\hspace*{5mm}
\includegraphics[width=5.4cm,angle=90,bbllx=80,bblly=80,bburx=530,bbury=730]{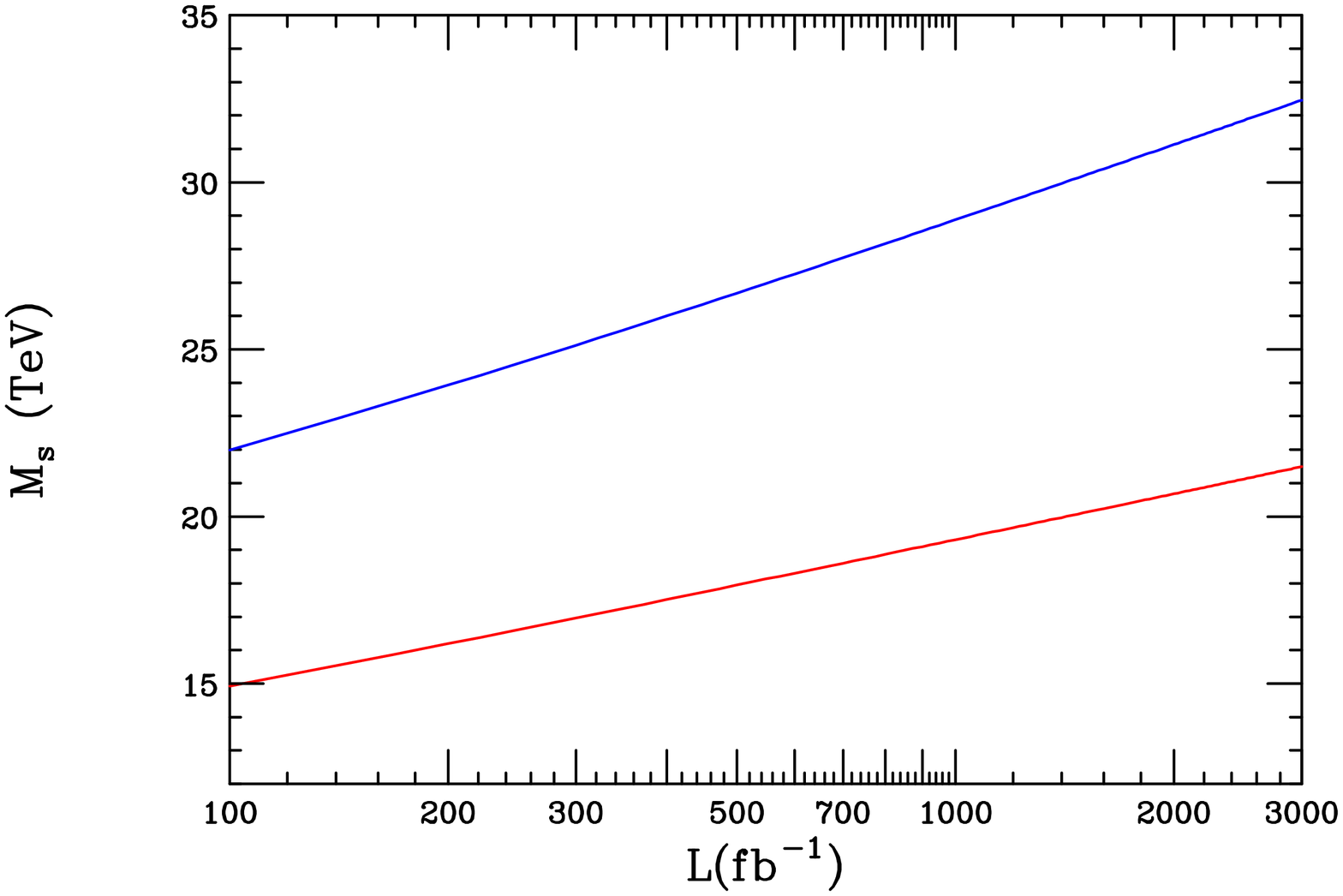}}
\vspace*{0.1cm}
\caption{(Left) Deviations in the cross section for  $A_{LR}$ 
for $b$-quarks at $\sqrt s$=5 TeV for $M_s=15$ TeV in the ADD model 
for an integrated 
luminosity of 1$~ab^{-1}$. The SM is represented by the histogram while the 
red and green data points show the ADD predictions with $\lambda=\pm 1$. In 
both plots $z=\cos \theta$.
(Right) Search reach for the ADD model scale $M_s$ at CLIC as a 
function of the integrated luminosity from the set of processes $e^+e^-\to 
f\bar f$ assuming $\sqrt s=3$(red) or 5(blue) TeV. Here $f=\mu,\tau,b,c$,  
$t$, etc~\protect \cite{rizzo}.}
\label{fig:ed1}
\end{figure}

\begin{figure}[htbp]
\centerline{
\includegraphics[width=5.4cm,angle=90,bbllx=80,bblly=80,bburx=530,bbury=730]{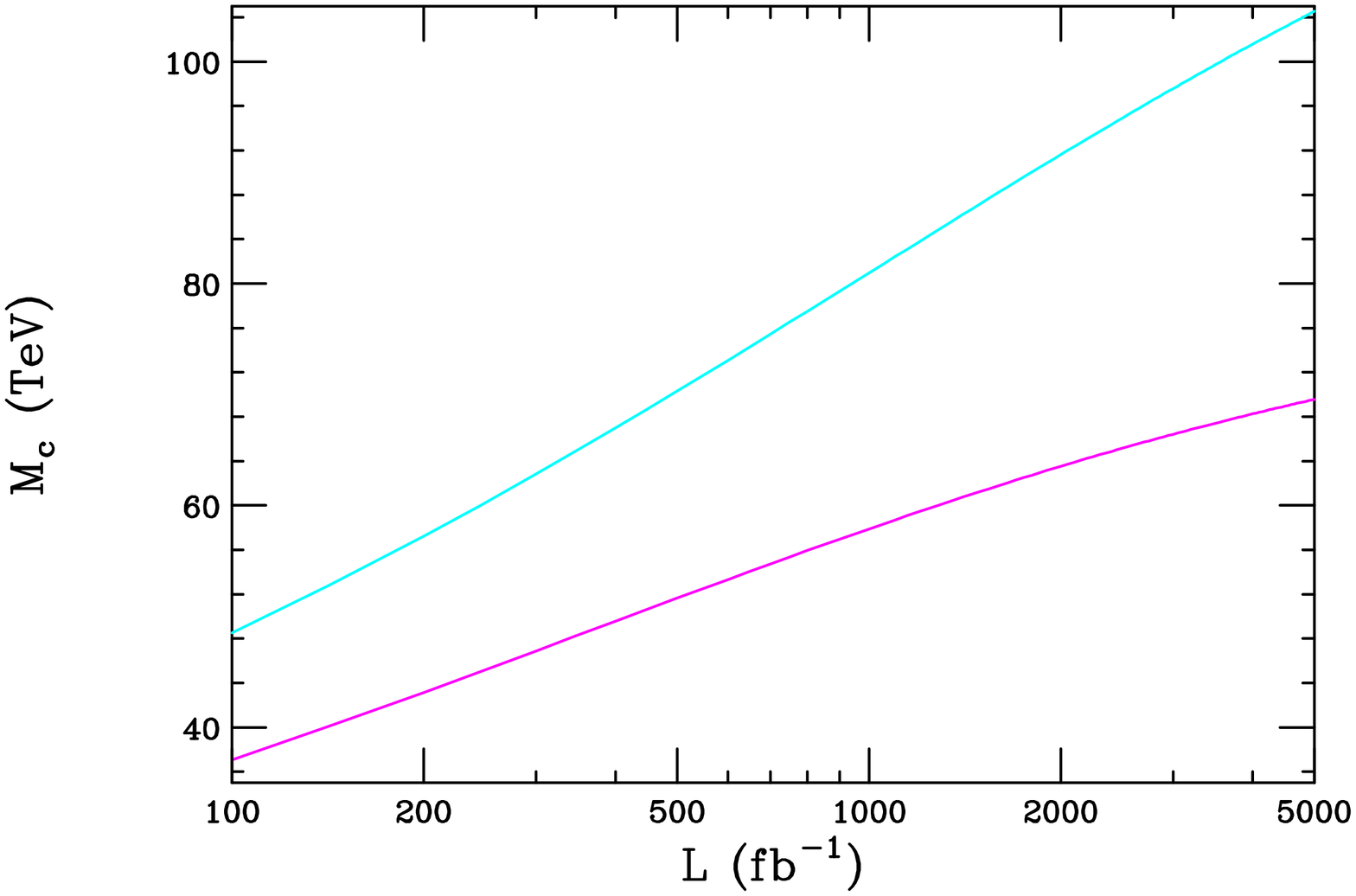}
\hspace*{5mm}
\includegraphics[width=5.4cm,angle=90,bbllx=80,bblly=80,bburx=530,bbury=730]{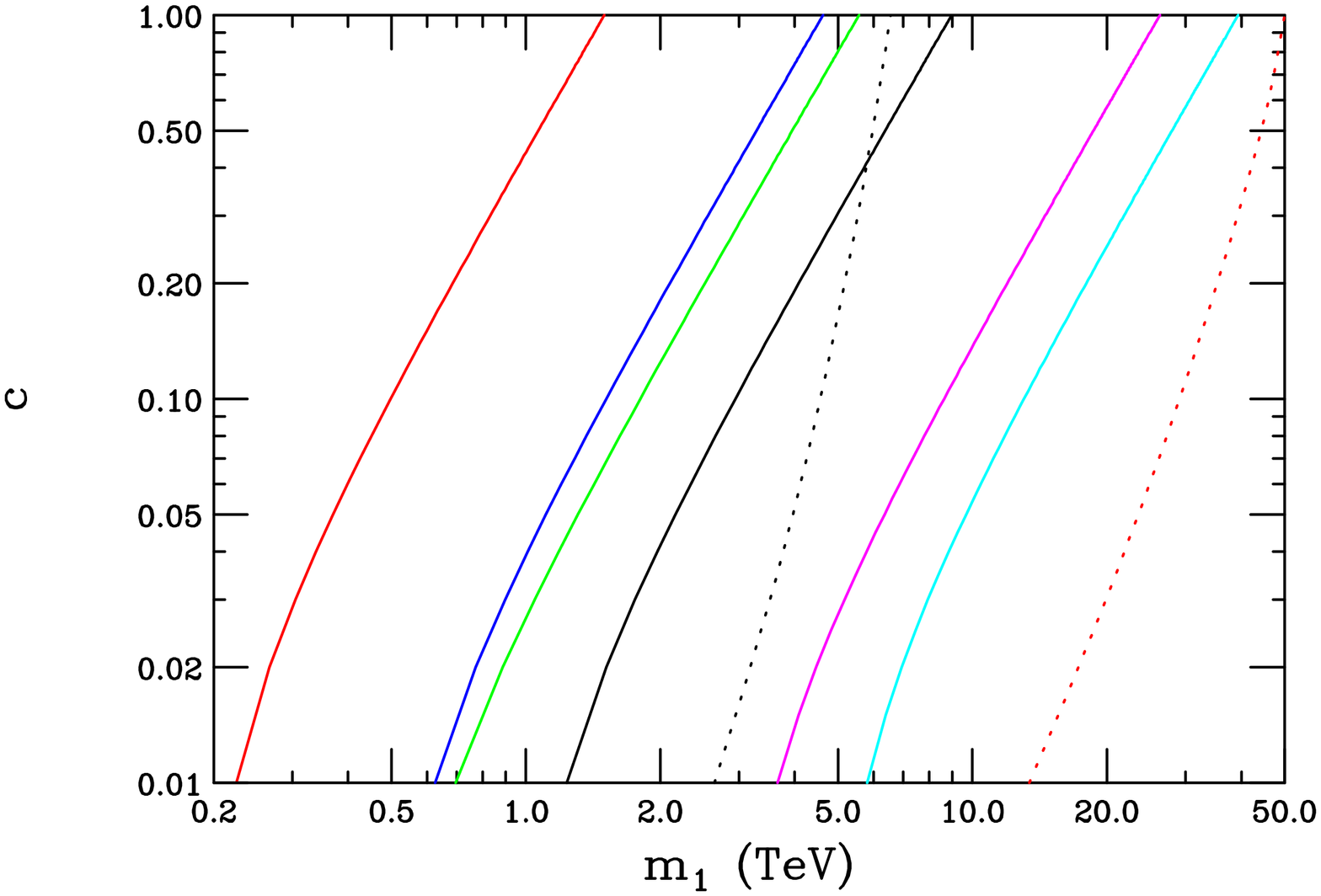}}
\vspace*{0.1cm}
\caption{(Left)
Corresponding reach for the compactification scale of the KK 
gauge bosons in the case of one extra dimension and all fermions localized 
at the same orbifold fixed point.
(Right)Indirect constraints from $e^+e^-$ colliders on the RS model 
parameter space with $c=k/\mpl$; the excluded region is to the left of the 
curves. From left to right the solid curves correspond to bounds from LEP II, 
a 500 GeV LC with 75 or 500 $fb^{-1}$ luminosity, a TeV machine with 200 
$fb^{-1}$, and a 3 or 5 TeV CLIC with 1 $ab^{-1}$. The dotted lines are the 
corresponding LHC (100 $fb^{-1}$) and $\sqrt s=175$ TeV VLHC(200 $fb^{-}$) 
direct search reaches~\protect \cite{rizzo}. }
\label{fig:ed2}
\end{figure}

\begin{figure}
\begin{center}
\epsfig{file=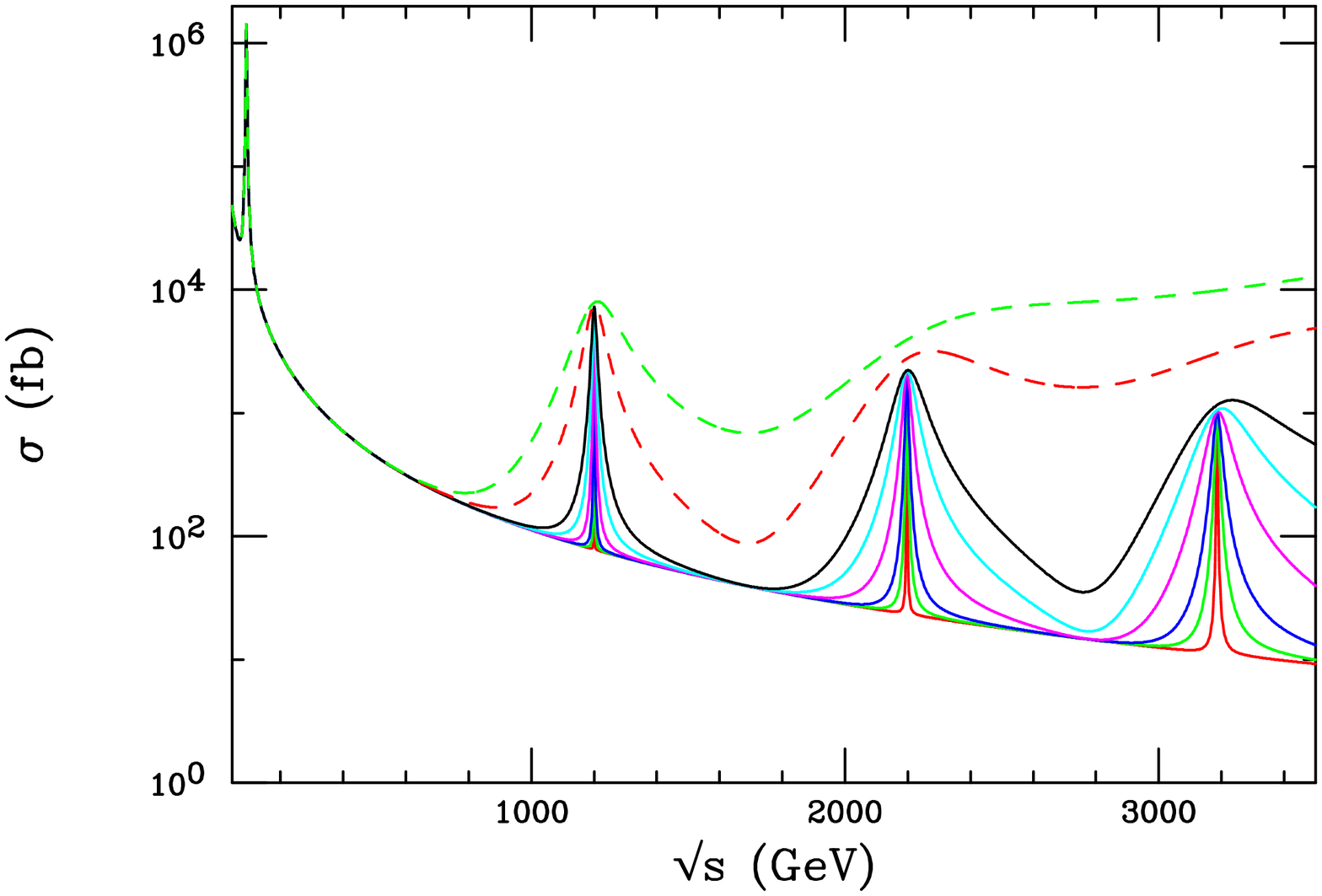,bbllx=90,bblly=90,bburx=550,bbury=720,width=7cm,angle=90,clip}
\epsfig{file=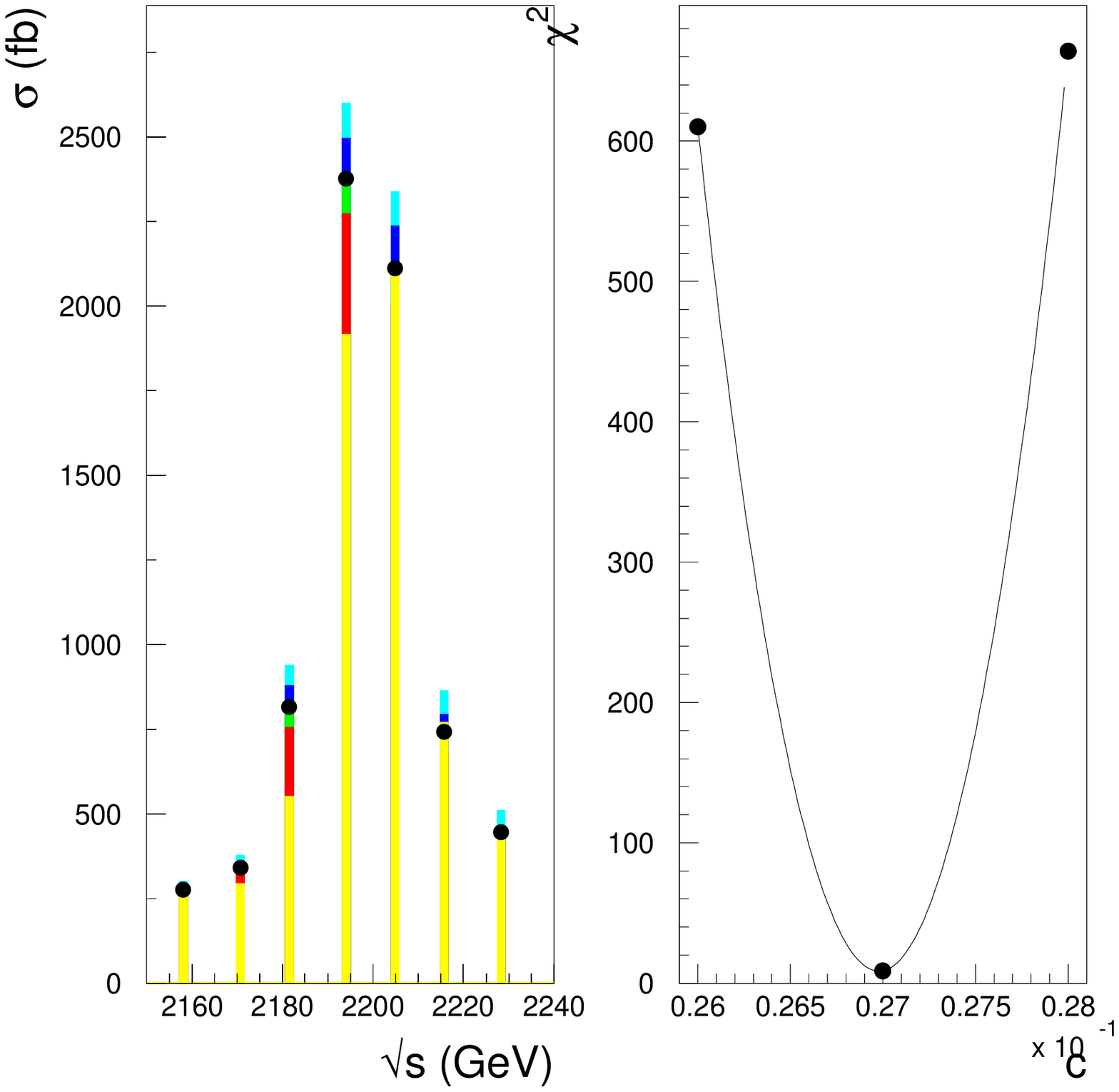,bbllx=30,bblly=160,bburx=540,bbury=650,width=7cm,clip}
\end{center}
\caption{
(Left) KK graviton excitations in the RS model 
produced in the process $e^+e^-\to \mu^+\mu^-$. From the most narrow to widest 
resonances the curves are for $c$ in the range 0.01 to 0.2. 
(Right) Scan of the resonance (7 points) and $\chi^2$ fit to the
measured spectrum for models with different values for 
$c$~\protect \cite{rizzo,deroeck2}.}
\label{fig:ed3}
\end{figure}

\begin{figure}[htb]   
\begin{center} 
\epsfig{file=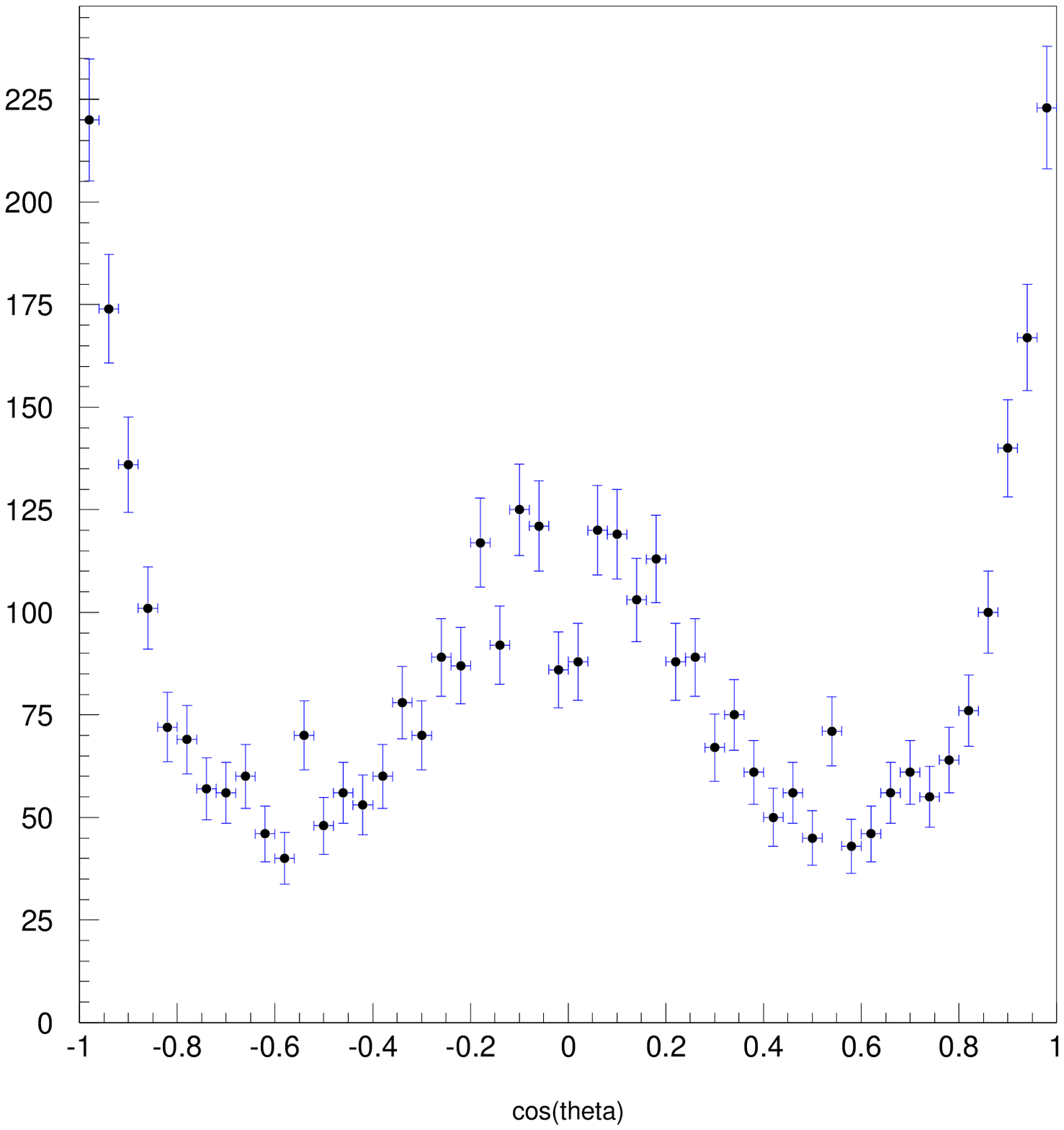,bbllx=0,bblly=0,bburx=540,bbury=540,width=8cm}
\epsfig{file=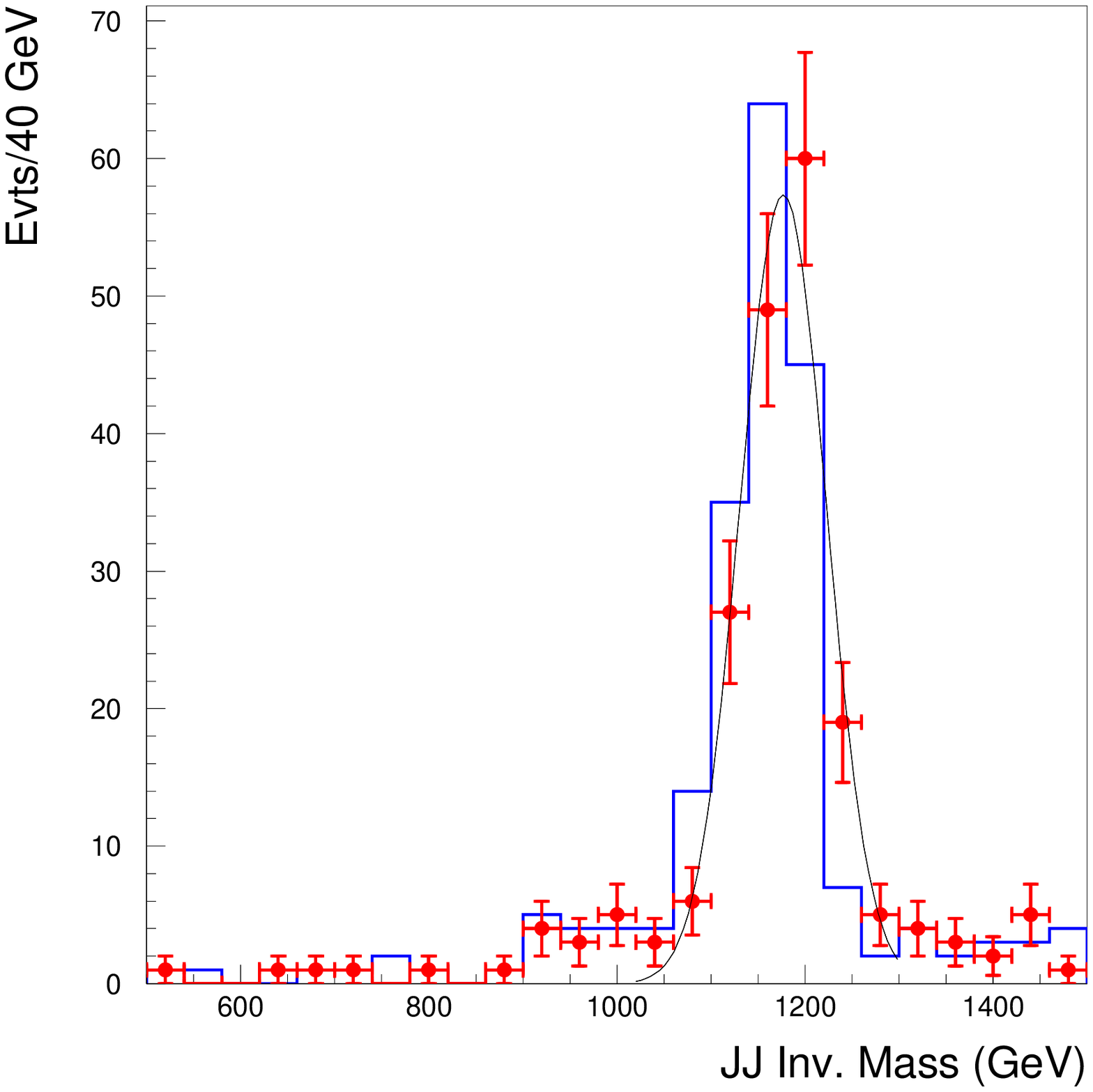,bbllx=0,bblly=0,bburx=565,bbury=565,width=8cm}
\end{center}
\caption{(Left) Angular distribution of produced muons from decay of a 
graviton. (Right) Reconstructed mass distribution of the $G_1$ graviton 
decaying into two jets~\protect \cite{deroeck2}.} 
\label{fig:ed5}

\end{figure}

\begin{figure}[htb]   
\begin{center} 
\includegraphics[height=7.0cm]{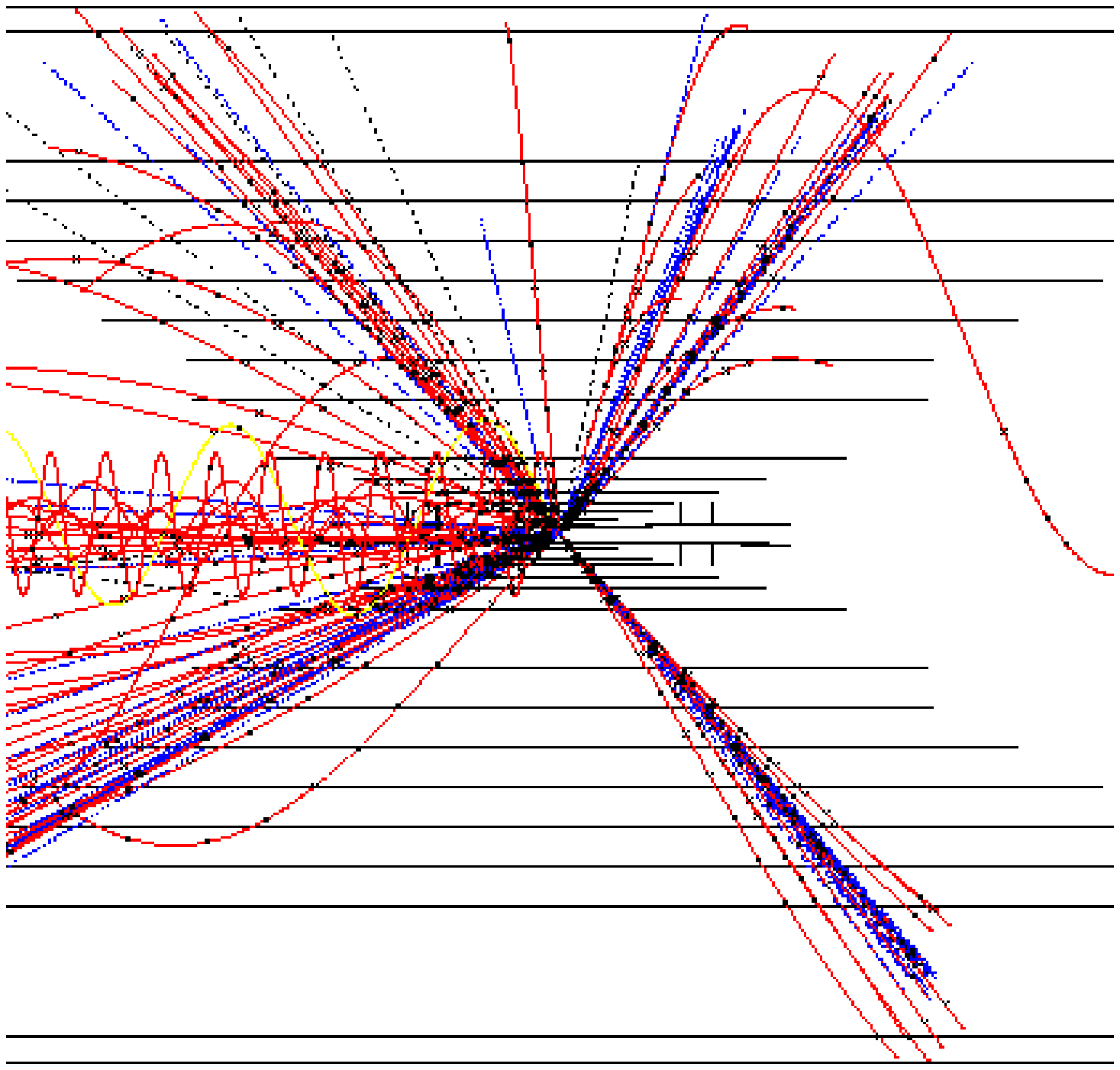}
\includegraphics[height=7.0cm]{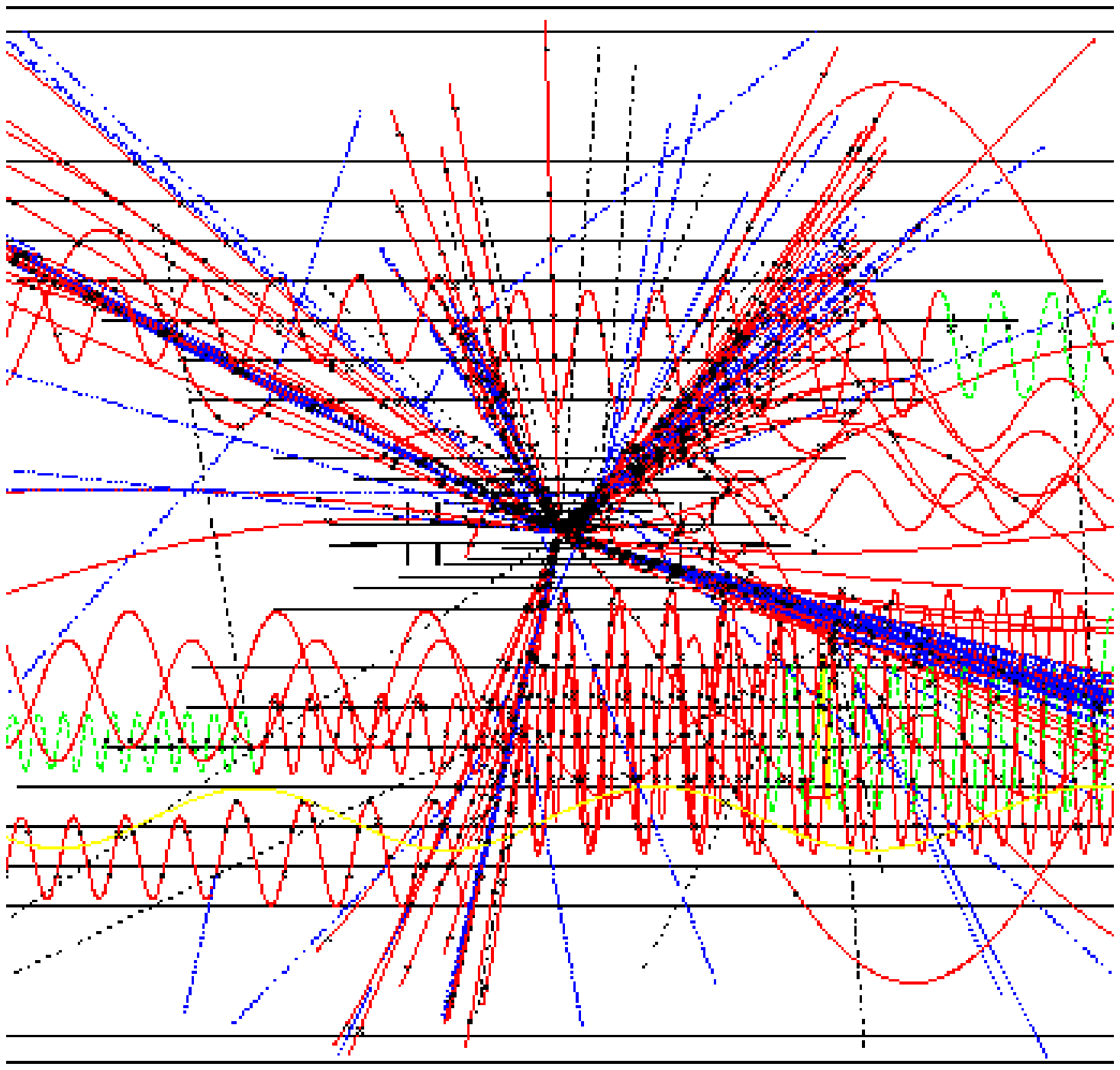}
\end{center}
\caption{Graviton resonance $G_3$ decaying into 
$G_1G_1 \rightarrow$ 4 jets~\protect \cite{deroeck2}.}
\label{fig:kk2}
\end{figure}

Extra dimension signatures at future colliders, particularly at
CLIC, 
 have been discussed
in detail in~\cite{rizzo}. 
There is a wide variety of models which can be basically
classified in three groups.
($i$) The first model is the so called large extra dimensions scenario of 
Arkani-Hamed, Dvali and Dimopoulos(ADD){\cite {add}}. This model predicts the 
emission and exchange of large Kaluza-Klein(KK) towers of gravitons that are 
finely-spaced in mass. The emitted gravitons appear as missing energy while 
the KK tower exchange leads to contact interaction-like dimension-8 operators. 
($ii$) A second possibility are models where the extra dimensions are of TeV 
scale in size. In these scenarios there are KK excitations of the SM gauge 
(and possibly other SM) fields with masses of order a TeV which can appear as 
resonances at 
colliders~\cite{anton}. ($iii$) A last class of models are those with warped 
extra dimensions, such as the Randall-Sundrum Model(RS){\cite {rs}}, which 
predict graviton resonances with both weak scale masses and couplings to 
matter. 

High energy $\ee$ colliders in the multi-TeV range with sufficient 
luminosity, such as CLIC, will be able to both directly and indirectly search 
for and/or make detailed studies of models in all three classes.

Figs.~\ref{fig:ed1} and \ref{fig:ed2} show the reach for indirect measurements
in the channel $e^+e^- \rightarrow f\overline{f}$
for a multi-TeV collider,  
from~\cite{rizzo}. The convention used in~\cite{hewetted} is adopted, 
expressing the contributions to spin-2 exchanges in terms of the scale $M_s$
and a sign $\lambda$.
Fig.~\ref{fig:ed1}a shows how such deviations from the SM can manifest
themselves 
at a 5 TeV CLIC  in case $M_s = 15 $ TeV, for the two signs of 
$\lambda$. For the evaluation of the search reach  
 several fermion final states
have been combined. The results on the search reach
versus the available luminosity  are shown in Fig.~\ref{fig:ed1}b.
Fig.~\ref{fig:ed2} shows search reaches for examples of the 
other two model classes (see ~\cite{rizzo}).
In summary a multi-TeV collider has a sensitivity to reach up to
20-80 TeV, depending on the model, from precision measurements.

Direct measurements can be made in the RS and TeV scale 
extra dimension models.
In the Randall-Sundrum model~\cite{rs} TeV scale graviton resonances are 
expected  in many channels. In its simplest version, with two branes and one 
extra dimension, and all of the SM fields remaining
 on the brane, the model has 
two fundamental parameters: the mass of the first resonance
and the parameter $c= k/M_{\overline{Pl}}$ which should be less than but
not too far away from unity. This parameter controls the effective coupling
strength of the graviton and thus the width of the resonances. 
A spectrum for $e^+e^-\rightarrow \mu^+\mu^-$ is given in 
Fig.~\ref{fig:ed3}a. Clearly the cross sections are huge and 
a putative signal cannot be missed 
by a LC with sufficient CMS energy.
This LC  will be like a KK resonance factory, up to the kinematic limit,
and can be used 
to measure the properties ($M, c$, branching ratios) 
and quantum numbers  (spin) of the KK
resonances.

The  signal for high energy resonance ($G_3$ at 3200 GeV)
has been selected via either
two muons or two $\gamma$'s with $E> 1200$ GeV and $|\cos\theta| <0.97$.
The background from overlaid two photon events is typically below 120 
mrad, thus mostly outside the signal region.
A scan similar to the one for the  $Z$ at LEP, 
folded with the CLIC luminosity spectrum,
was made for an integrated luminosity of  1 ab$^{-1}$, and results
in a precision to  determine $c$ to  0.2\% and $M$ to better than 
0.1\%~\cite{deroeck2}.
An example of such a scan 
is shown in Fig.~\ref{fig:ed3}b.

The graviton is a spin-two object. Fig.~\ref{fig:ed5} shows the 
decay angle of the  fermions from $G\rightarrow \mu\mu$ 
for the lightest graviton,  for 1  ab$^{-1}$ of data,
using signal events and including machine background.
The spin-two nature of the resonance is clearly visible.
Another property of these gravitions is 
{$BR(G \rightarrow \gamma\gamma)/BR(G \rightarrow \mu\mu)$ = 2}.
These events can be easily selected by cuts on isolated muons and photons, 
even in the presence of machine background~\cite{deroeck2}. Hence 
the ratio $BR(G\rightarrow \gamma\gamma)/
BR(G\rightarrow
\mu\mu)$ can be measured with a precision  better than 1\%.

Furthermore,
if 
the energy of the collider is large enough  to see the first 
three graviton resonances, there 
is the intriguing 
opportunity
to measure the graviton 
self coupling in 
$G\rightarrow GG $ decay ($\sim$ 15\% of decays)~\cite{rizzo3}.
The dominant decay mode is $G \rightarrow$ gluon-gluon 
into jets. A few examples of  events decaying into 4 jets  
are  shown in Fig.~\ref{fig:kk2}.
Fig.~\ref{fig:ed5} shows the reconstruction
of the $G_{1}$ mass from 2 jets 
from $G_{3} \rightarrow G_{1} G_{1}$ decay.
The histogram contains  no background, while  the points
 contain 10 bunch crossings of $\gamma\gamma$ background  overlaid.
The $G_1$ graviton signal 
can be easily reconstructed in the decay of the $G_3$ graviton.

If KK resonances are produced in the TeV region, then
 CLIC can be used to precisely determine the shape,
mass,  spin, and branching ratios of the objects.  With such information the objects 
can be unambiguously identified as gravitons.

\begin{figure}[htbp]
\begin{center}
\epsfig{file=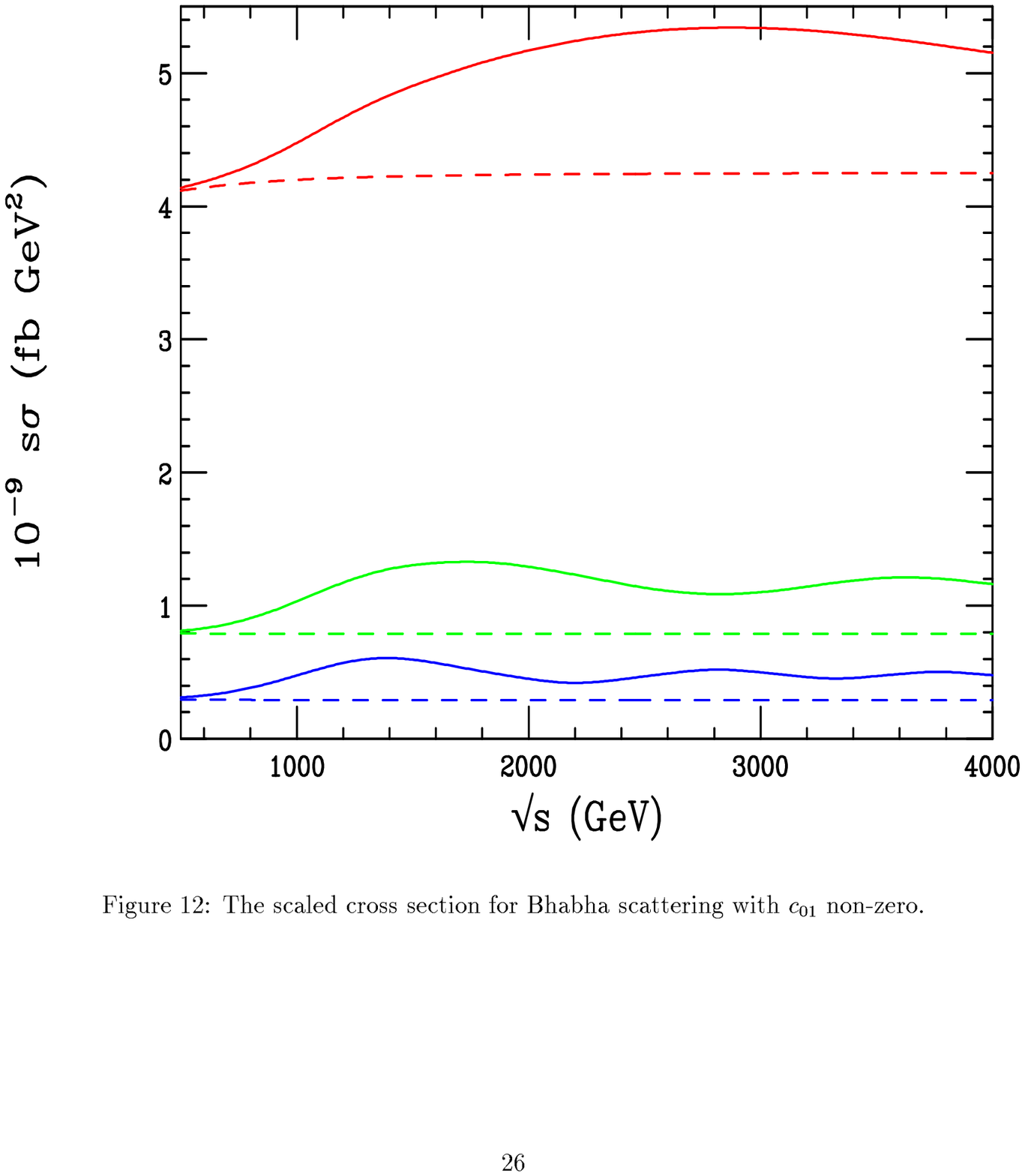,bbllx=62,bblly=227,bburx=555,bbury=645,width=10cm,
height=7.5cm,clip}
\end{center}
\caption{The scaled cross section for Bhabha scattering including
$\cos \theta$ cuts on the scattered electrons:
(from top to bottom) $\cos\theta < 0.9/0.7/0.5$,
for $\Lambda_{NC} = 500 $ GeV. Dashed curves are the SM 
production~\protect \cite{hewett}.}
\label{fig:nc}
\end{figure}

\begin{figure}[htbp]
\begin{center}
\includegraphics[height=7.0cm]{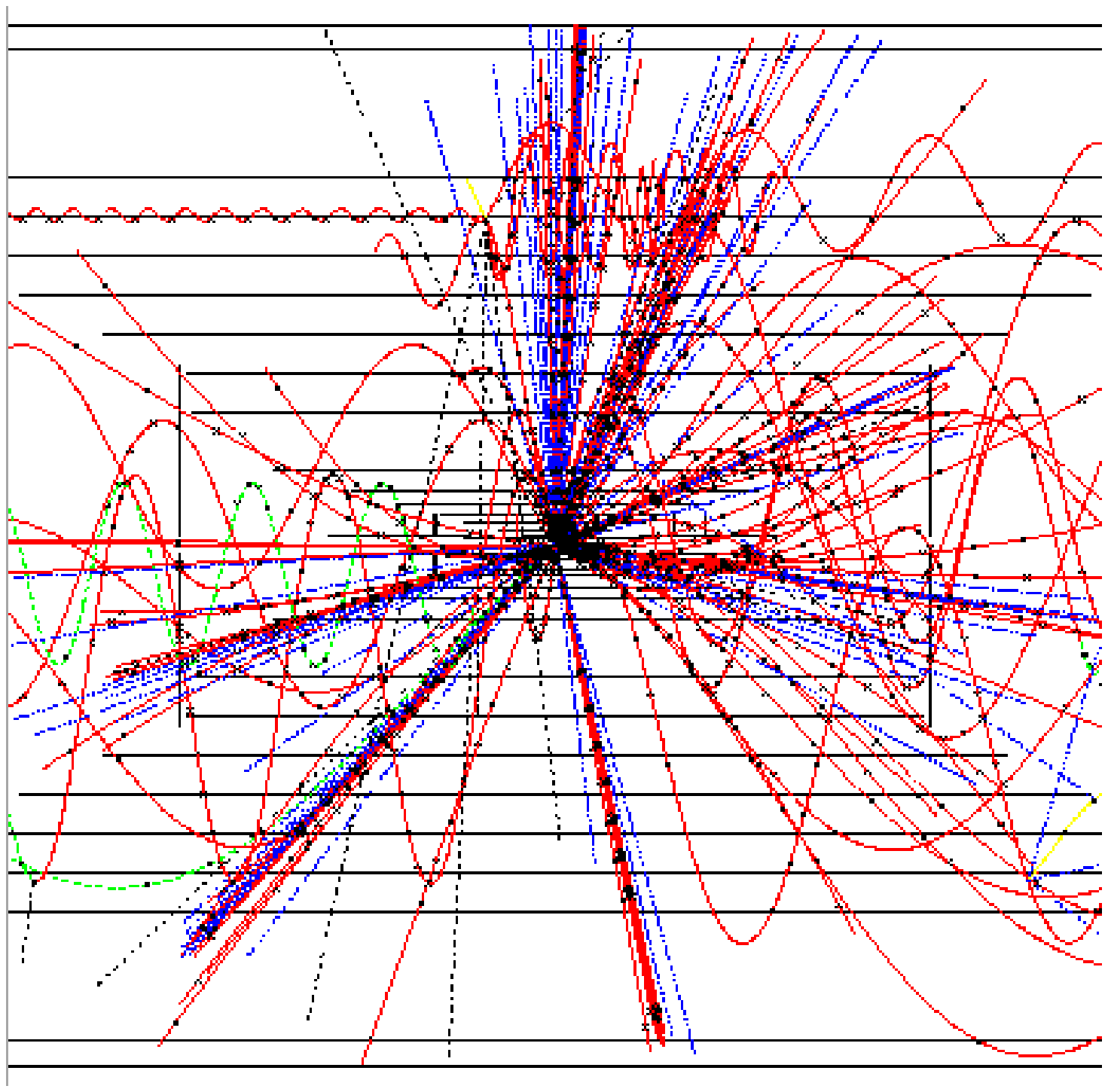}
\includegraphics[height=7.0cm]{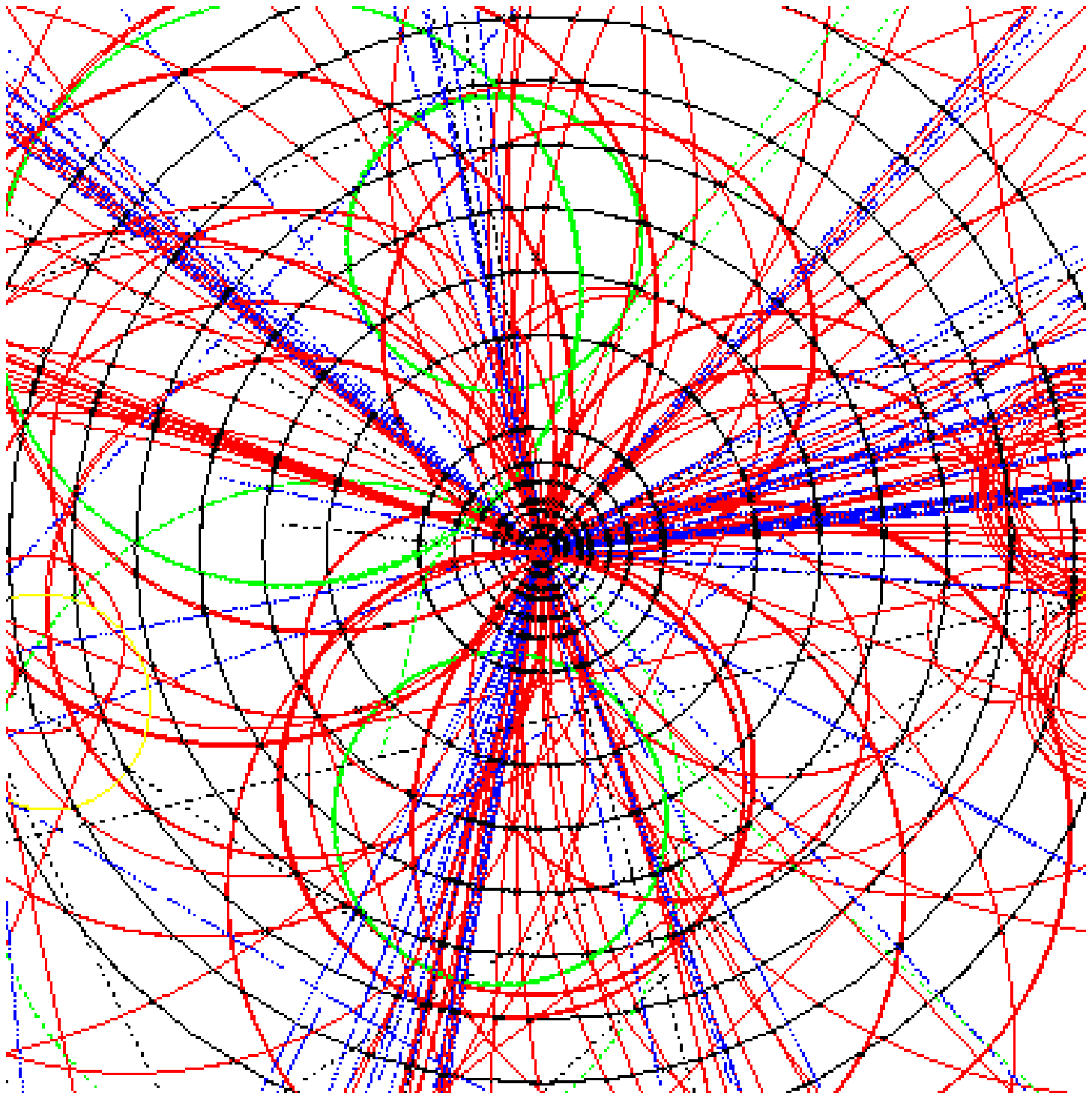}
\end{center}
\caption{Black hole production in the CLIC detector.}
\label{fig:bh}
\end{figure}

\section{Non Commutative Theories}

Recently theoretical results have demonstrated that non-commutative 
(NC) geometries
naturally appear within the context of string/M theories~\cite{hewett}.
 In such theories the conventional space-time coordinates,
which are represented by operators, no longer commute, due to 
  a preferred direction in space, and Lorentz invariance is 
violated. The effect can be parameterized by a scale $\Lambda_{NC}$, 
which characterizes the threshold at which NC effects become important.
The most likely value of $\Lambda_{NC} $ is close to the Planck scale,
and with 
the recent ideas of extra dimensions in which the Planck scale might be as
low as a few TeV, there exists 
the possibility of observing 
such  stringy effects at colliders.

A consequence of a NC quantum field theory is that QED
takes on a non-abelian nature due to 
new
  3 and 4-point photon couplings. 
This will have striking effects on Bhabha and 
M$\o$ller scattering, and on the reaction 
$\gamma\gamma\rightarrow \gamma\gamma$.
The non-abelian nature of QED can be tested  
 by looking for unexpected structure in kinematic variable distributions caused by
$s$- and $t$- channel interference.
For example, Fig.~\ref{fig:nc} shows the energy dependence of the  
scaled Bhabha cross section as function
of the CMS energy, for different $\cos\theta$ selections for the 
scattered electrons.
The variation results from the momentum dependent phase factors 
present at the QED 
vertices. The effects can be large and modulations of order 
20\%-30\% are
expected for this choice of model parameters~\cite{hewett}.

The sensitivity to $\Lambda_{NC}$ for a 3 and 5 TeV collider has been 
studied in detail in~\cite{hewett}. 
For the reaction $e^+e^-\rightarrow \gamma\gamma$
the search reach (95\%) for 1 ab$^{-1}$ is $\Lambda_{NC} \sim
2.5-3.5 $ (3 TeV) and $3.8-5.0$ (5 TeV).
For the reaction $e^+e^-\rightarrow e^+e^-$
the search reach (95\%) for 1 ab$^{-1}$ corresponds
to $\Lambda_{NC} \sim
4.5-5.0 $ (3 TeV) and $6.6-7.2$ (5 TeV), and shows  a larger sensitivity.

Hence NC effects, if present at the TeV scale, will produce large 
characteristic effects
in the data at a multi-TeV collider.

\section{Black Hole Production}

Another idea to emerge from all the recent work on extra dimensions is the laboratory production of black holes. 
If the fundamental Planck scale is in the TeV range
then a natural consequence would
be the possibility of  
black hole (BH) production in the multi-TeV range, above the 
Planck scale.
The cross section is  expected to be very large:
$\sigma = \pi R_s^2 \sim 1 $ TeV$^{-2}\sim O(100) $ pb,
where $R_S$ is the  Schwarzschild Radius.
If $\sqrt{s}_{e^+e^-} >M_{BH}>M_{\rm Planck}$ then the 
collider becomes a  black hole factory.
The lifetime of such a black hole is of order 
 $ \sim 10^{-25}-10^{-27}$ sec, and hence the black hole will evaporate before
it could possibly `attack' any detector material. Hence these 
mini black holes are not dangerous for mankind.  Furthermore, if black hole production can take place in 
multi-TeV $\ee$ collisions, then cosmic ray neutrinos
with energies in excess of $10^{3}$ TeV have been producing mini black holes
in collisions with earth's atmosphere throughout its history~\cite{Feng:2001ib}.

The decay of a black hole can be very complex and involve several
stages~\cite{thomas,landsberg}. If the dominating mode is 
 Hawking radiation then all 
particles (quarks, gluons, gauge bosons, leptons) are expected to 
be produced democratically, with e.g. a ratio 1/5 
between leptonic and hadronic activity. 
The multiplicity is expected to be large.
The production and decay process
have been included in the PYTHIA generator\cite{landsberg}.
Fig.~\ref{fig:bh} shows two black hole events produced in a detector
at CLIC, leading to spectacular multi-jet and lepton/photon signals.
If this scenario is realized in Nature black holes will be produced
at high rates at the LHC. CLIC can be very instrumental 
in providing 
precise measurements.  For example, it could be used to test Hawking radiation and 
extract the number of underlying extra dimensions.

The large production cross section, low backgrounds and little missing 
energy would make BH production and decay a perfect laboratory to study 
strings and quantum gravity.

Besides the black holes other 
 novel ideas include:

\begin{itemize}

\item String quantum gravity effects in $e^+e^- \rightarrow e^+e^-$ and 
$\gamma\gamma$~\cite{peskin}

\item Brane fluctuation modes (Nylons?)~\cite{lykken}

\item Split Fermions in Extra Dimension~\cite{arkani}

\item Extra Dimensions \& Two Supersymmetries~\cite{Hall}

\end{itemize}

If any of these scenarios  occurs in Nature
a multi-TeV $e^+e^-$ collider promises to be extremely instrumental
and useful to  
explore this new physics domain.

\begin{table}[htb]
\caption{Measurements at CLIC (5 TeV / 1 ab$^{-1}$, unless otherwise
stated).}
\begin{center}
\begin{tabular}{|l|l|}
\hline
Higgs (Light) & $g_{HHH}$ to $ \sim 7-10\%$ (5 ab$^{-1}$/ 3 TeV) \\
Higgs (Light) & $g_{H\mu\mu}$ to $ \sim 4-10\%$ (5 ab$^{-1}$/ 3 TeV) \\
Higgs (Heavy) & 2.0 TeV $(e^+e^-)$  3.5 TeV $(\gamma\gamma)$ \\
squarks &  2.5 TeV\\
sleptons & 2.5 TeV\\
Z' (direct) & 5 TeV\\
Z' (indirect) & 20-30 TeV\\
$l^*,q^*$ & 5 TeV\\
TGC (95\%): $\Delta \lambda_{\gamma}$ & 0.00008\\
TGC (95\%): $\Delta \kappa_{\gamma}$ & 0.00005\\
$\Lambda$ compos. & 400 TeV\\
$W_{L}W_{L}$ 
& $> 5$ TeV \\
ED (ADD) & 30 TeV  ($e^+e^-$              )\\
  & 55 TeV  ($\gamma\gamma$              )\\
ED (RS) & 18 TeV (c=0.2)\\
ED (TeV$^{-1}$) & 80 TeV\\
Resonances & $\delta M/M, \delta \Gamma/\Gamma \sim 10^{-3}$\\
Black Holes & 5 TeV \\
\hline
\end{tabular}
\end{center}
\label{tab:sum}
\end{table}

\section{Summary}

Linear $e^+e^-$ colliders operating in the multi-TeV energy range
are likely to be based on the CLIC two-beam acceleration concept.
To achieve a large luminosity, such an accelerator would need to
operate in the high beamstrahlung region, rendering 
experimentation at such a collider more challenging. Studies so far
indicate that this is not a substantial handicap, and the precision
physics expected from an $e^+e^-$ collider will be possible.
  
 The two-beam  accelerator technology is not yet available today
for use at a large scale collider. 
 R\&D on this technology will continue until 2006 at least, 
after which -- if no bad surprises emerge-- one can plan 
for a full technical design of such a collider.

From the physics program side, 
a multi-TeV collider has a large potential to push back the high energy
horizon further, up to scales of 1000 TeV, where 
-- if the Higgs is light-- new physics can no longer hide from
experiment.
If no new scale is found by then we have to revise our 
understanding of Nature.

A multi-TeV collider with high luminosity can be used for precision 
measurements in the Higgs sector.  It can precisely measure
the masses and couplings of  heavy sparticles, 
thereby completing the SUSY spectrum.
If extra dimensions or even black holes pop up in the 
multi-TeV range, such a collider will
be a precision instrument to study quantum gravity in the laboratory.

The physics reach, as envisioned today, for a multi-TeV collider is 
summarized in Table~\ref{tab:sum}. In short
a collider with  $\sqrt{s} \simeq$~3-5~TeV is expected to break new grounds,
beyond the LHC and a TeV class LC.

\begin{acknowledgments}
We are grateful to the participants of the Snowmass E3-S2 subgroup on 
multi-TeV
$e^+e^-$ colliders,  and to the CLIC physics study group, in particular
T. Rizzo, J. Hewett, M. Battaglia, and D. Schulte.
\end{acknowledgments}


\end{document}